\documentclass[a4paper]{aa}
\usepackage{lscape}
\usepackage{gensymb}
\usepackage{txfonts}
\usepackage{natbib}
\bibpunct{(}{)}{;}{a}{}{,} 

\usepackage{pdfpages}
\usepackage{graphicx}
\usepackage{natbib}
\usepackage{longtable,lscape}
\usepackage{amssymb}
\usepackage{mathrsfs}
\usepackage{amsfonts}
\usepackage{url}
\usepackage{color}

\usepackage[flushleft]{threeparttable}

\def\Planck{\textit{Planck}}

\def\Planck{\textit{Planck}}
\def\deg{\ifmmode^\circ\else$^\circ$\fi}
\def\pdeg{\ifmmode $\setbox0=\hbox{$^{\circ}$}\rlap{\hskip.11\wd0 .}$^{\circ}
          \else \setbox0=\hbox{$^{\circ}$}\rlap{\hskip.11\wd0 .}$^{\circ}$\fi}
\def\arcs{\ifmmode {^{\scriptstyle\prime\prime}}
          \else $^{\scriptstyle\prime\prime}$\fi}
\def\arcm{\ifmmode {^{\scriptstyle\prime}}
          \else $^{\scriptstyle\prime}$\fi}
\newdimen\sa  \newdimen\sb
\def\parcs{\sa=.07em \sb=.03em
     \ifmmode \hbox{\rlap{.}}^{\scriptstyle\prime\kern -\sb\prime}\hbox{\kern -\sa}
     \else \rlap{.}$^{\scriptstyle\prime\kern -\sb\prime}$\kern -\sa\fi}
\def\parcm{\sa=.08em \sb=.03em
     \ifmmode \hbox{\rlap{.}\kern\sa}^{\scriptstyle\prime}\hbox{\kern-\sb}
     \else \rlap{.}\kern\sa$^{\scriptstyle\prime}$\kern-\sb\fi}
\def\mo{\ifmmode^{-1}\else$^{-1}$\fi}

\begin{document}

\title{Optical validation and characterization of \Planck\ PSZ2 sources at
the Canary Islands observatories. I. First year of LP15 observations}

\titlerunning{Optical validation of \Planck\ PSZ2. I. }

\author{A.~Streblyanska \inst{1,2} \and
A.~Aguado-Barahona \inst{1,2} \and 
A.~Ferragamo\inst{1,2} \and
R. Barrena \inst{1,2}  \and
J.~A.~Rubi\~{n}o-Mart\'{\i}n \inst{1,2} \and
D.~Tramonte \inst{3,1,2} \and
R.~T.~Genova-Santos \inst{1,2} \and
H.~Lietzen \inst{4}
}
 
\institute{Instituto de Astrof\'{\i}sica de Canarias, C/ V\'{\i}a L\'{a}ctea s/n, E-38205 La Laguna, Tenerife, Spain\\ 
     \email{alina@iac.es} \and
 Universidad de La Laguna (ULL), Dept. Astrof\'{\i}sica, E-38203 La Laguna, Tenerife, Spain   \and
 School of Chemistry and Physics, University of KwaZulu-Natal, Westville Campus, Private Bag X54001, Durban, 4000, South Africa \and
 Tartu Observatory, University of Tartu, Observatooriumi 1, 61602, Toravere, Estonia
}

\date{Accepted 26/05/2019}

\authorrunning{Streblyanska et al. (2019)}

\abstract{The second catalogue of \Planck\ Sunyaev-Zeldovich (SZ) sources, hereafter PSZ2,
is the largest sample of galaxy clusters selected through their SZ signature
in the full sky. At the time of publication, 21\,\% of these objects had no
known counterpart at other wavelengths. Using telescopes at the Canary Island
observatories, we conducted the long-term observational programme
128-MULTIPLE-16/15B (hereafter LP15), a large and complete optical follow-up
campaign of all the unidentified PSZ2 sources in the northern sky, with
declinations above $-15^{\circ}$ and no correspondence in the first Planck
catalogue PSZ1.}
{The main aim of LP15 is to validate all those SZ cluster
candidates, and to contribute to the characterization of the actual purity and
completeness of full \Planck\ SZ sample. In this paper, we describe the full
programme and present the results of the first year of observations.}
{The LP15 programme was awarded 44 observing nights, spread over two years in
three telescopes at the Roque de los Muchachos Observatory. The full LP15
sample comprises 190 previously unidentified PSZ2 sources. For each target, we
performed deep optical imaging and spectroscopy. Our validation procedure
combined this optical information with SZ emission as traced by the publicly
available \Planck\ Compton $y$-maps. The final classification of the new
galaxy clusters as optical counterparts of the SZ signal is established
according to several quantitative criteria based on the redshift, velocity
dispersion and richness of the clusters.}
{This paper presents the detailed study of 106 objects out of the LP15 sample,
corresponding to all the observations carried out during the first year of the
programme. We confirmed the optical counterpart for  41 new PSZ2 sources, being
31 of them validated using also velocity dispersion based on our spectroscopic 
information. This is the largest dataset of newly confirmed PSZ2 sources without 
any previous optical information. All the confirmed counterparts are rich structures 
(i.e. they show high velocity dispersion), and are well aligned with the nominal
\Planck\ coordinates (i.e., $\sim 70$\,\% of them are located at less than
3$\arcmin$ distance). In total, 65 SZ sources are classified as unconfirmed,
being 57 of them due to the absence of an optical over-density, and 8 of them
due to a weak association with the observed SZ decrement. Most of the sources
with no optical counterpart are located at low galactic latitudes and present
strong galactic cirrus in the optical images, being the dust contamination the
most probable explanation for their inclusion in the PSZ2 catalogue.}{}


\maketitle


\section{Introduction}

The study of the anisotropies in the cosmic microwave background (CMB) has
become one of the most powerful tools in the field of cosmology and physics of
the early Universe. A large amount of new observational data collected from
different facilities and observatories allows testing and validating our
cosmological model down to per cent accuracy \citep{planck2015_xiii,
planck2018_vi}. However, further progress is still needed in the measurement
of the cosmological parameters if we want to solve some of the tensions that
appear when comparing the CMB constraints on some parameters with those coming
from other astrophysical probes \citep{planck2018_vi}. For example, the
measurement of the mass density of the Universe $\Omega_{\rm m}$ or the
normalization of the matter power spectrum $\sigma_8$ remain important and
challenging problems \citep{planck2015_a24}.

Progress in this topic might come from detailed studies on clusters of
galaxies. Massive clusters are particularly sensitive to the cosmology
\citep[e.g.,][]{planck2013-p15}. However, due to its rareness, only large-volume
surveys could present big enough samples for detailed investigations.

In the last decade, the Sunyaev-Zeldovich (SZ) effect \citep{sz1972} has become
a powerful technique in cosmology as an instrument to detect such massive
clusters. This effect produces a spectral distortion of the cosmic microwave
background (CMB) generated by the inverse Compton interaction between the CMB
photons and the hot intracluster gas of electrons.  Today, thanks to the
\Planck\footnote{\Planck\ \url{http://www.esa.int/Planck} is a project of the
European Space Agency (ESA) with instruments provided by two scientific
consortia funded by ESA member states and led by Principal Investigators from
France and Italy, telescope reflectors provided through a collaboration
between ESA and a scientific consortium led and funded by Denmark, and
additional contributions from NASA (USA).} satellite we were able for the
first time to detect galaxy clusters via the SZ effect in a full-sky survey
\citep{planck2013-p05a,plPSZ2}.

During the first 1.5 years of the mission, several catalogues have been
released: the early ESZ catalogue \citep{planck2011}, and the first official
\Planck\ SZ catalogue \citep[][hereafter, PSZ1]{planck2013-p05a}. In 2015, the
second and final catalogue, based on the full mission data, was released
\citep[][hereafter, PSZ2]{plPSZ2}. It contains 1653 sources (in comparison with
1227 clusters presented in PSZ1), and is the largest SZ-selected sample of
galaxy clusters candidates today.

Despite the detailed validation process carried out by the
\Planck\ collaboration, there was a significant fraction of the sample (559
objects at the time of the PSZ2 publication) with no known
counterparts. Moreover, some of the SZ sources, especially those with a low
significance SZ detection, may correspond to spurious enhancements of the
SZ signal due to the galactic dust contribution at the \Planck\ higher
frequencies \citep[e.g.][]{pl48}. A systematic follow-up study at other
wavelengths might help to disentangle this problem. In addition, being the SZ
surface brightness independent on redshift, the SZ flux alone cannot provide
either the distance or the dynamical mass information of the clusters.

Thus, in order to make the SZ catalogues appropriate for cosmological studies,
in 2010--2018 the \Planck\ Collaboration and several individual research groups
performed a few extensive follow-up programmes dedicated to confirm SZ sources
from those catalogues, using ground-based and space facilities
\citep[e.g.][]{planck4, planck2014-XXVI, liu2015, burg16, bur18,
  boada18}. Telescopes at Canary Islands Observatories have actively participated
in these validation and characterization efforts of \Planck\ catalogues
\citep[][]{planck2015-XXXVI, barrena18}. The first two catalogues, ESZ and PSZ1,
have been partially observed and validated through the International Time
Programme\footnote{ITP: \url{http://www.iac.es/eno.php?op1=5&op2=13&lang=en}}
{\tt ITP12-2} and {\tt ITP13-08} \citep[e.g.][]{planck2015-XXXVI, barrena18,
  rub18}. As a result, we retrieved physical properties and confirmed almost 200
previously unknown clusters with $0.08 < z < 0.85$. This validation programme
provided also important information for cosmological studies with clusters. For
example, it was used to constrain the bias between SZ mass and dynamical mass
\citep{fer19}.

The new PSZ2 catalogue\footnote{The PSZ2 catalogue (FITS format) and a
detailed description of its content can be found at the \Planck\ 2015 Release
Explanatory Supplement \url{https://wiki.cosmos.esa.int/planckpla2015/}. }, as
a final and uniformly selected sample, provides new galaxy cluster candidates
which could be used to constrain cosmological parameters. With this aim, and
taking into account the impact of our previous observational efforts for PSZ1,
the new long-term follow-up programme has been launched in order to validate the
second and final \Planck\ catalogue. This programme is the largest and the most
complete optical follow-up campaign of unconfirmed PSZ2 sources in the northern
sky.

The current paper describes the rationality and strategy assumed for the {\tt
  LP15} follow-up campaign, together with the results from the first year of
observations. The results of the second year observations, and the final
conclusions of the programme, will be presented in \citet{alej19}, in
preparation.

This paper is organized as follows. Section \ref{sec:int} describes the
\Planck\ PSZ2 catalogue and present motivation and strategy of our long-term
programme, observation and data reduction. Section \ref{sec:new} shows the
methodology used for new cluster identification.  In Section~\ref{sec:disc}, we
discuss the results of the PSZ2 characterization exposed in this work, including
a detailed description of some particular SZ targets (multiple detections,
presence of gravitational arcs etc). Finally, in Section~\ref{sec:conclusions}
we present the conclusions.

We adopt $\Lambda$CDM cosmology with $\Omega_{\rm m}$= 0.3075, $\Omega_\Lambda$=
0.691 and H$_0$= 67.74\,km\,s$^{-1}$\,Mpc$^{-1}$.

\begin{table*} 
\begin{center}
\caption[]{Summary information for the 2-years long-term
programme {\tt LP15}. Columns 4 to 6 show the field of view, the pixel scale
and the resolution of each telescope used (either imaging or spectroscopic
mode). Column 7 shows the total number of awarded nights per telescope. The
last two columns present the total number of observed SZ clusters (imaging and
spectroscopy), with the separated information for the first and second year of
the program.}
\label{tab:telescopes}
\begin{tabular}[h]{ccccccccccccc}
    \hline \hline
Telescope & Aperture [m] & Instrument &  FoV & Pixel Scale ["] & Resolution & N$_{\textrm{nights}}$ &  N$_{\textrm{ima}}$(y1/y2)  &  N$_{\textrm{spec}}$(y1/y2) \\
 \hline \hline
  INT&  2.5 &   WFC    & $34\arcmin \times 34\arcmin$    & 0.33 & --     &  21 &       210 (102/108) & --  \cr
  TNG&  3.5 &  DOLORES &  $8.\arcmin6 \times 8.\arcmin6$ & 0.252& $R=600$&  13 &      --            & 20 (14/6) \cr
  GTC& 10.4 &  OSIRIS  &  $7.\arcmin8 \times 7.\arcmin 8$& 0.254& $R=500$&  80h (=10 nights) & -- & 44 (23/21) \cr
\hline   
  \end{tabular}
\end{center}
\small 
\end{table*}

\section{LP15 Optical follow-up campaign}
\label{sec:int}
\subsection{The \Planck\ PSZ2 cluster sample}
\label{sec:sample}

The PSZ2 catalogue has been produced from the data of the full-mission survey of
\Planck\, corresponding to 29 months of observations. It contains 1653
detections; 937 sources are common to PSZ1, while 716 are unique. 291 sources
from the PSZ1 catalogue are not anymore in PSZ2.

The methodology used to construct the PSZ2 is an extended and refined version of
the PSZ1 one \citep{planck2013-p05a}. The PSZ2 cluster candidates, similarly to
PSZ1, are blindly selected using three different detection algorithms
\citep{plPSZ2}. Two of the algorithms (MMF1 and MMF3) are based on the same
Matched Multi-filters technique. The third one (PowellSnakes or PwS) is based on
a fast Bayesian approach to discrete object detection. The catalogue contains
all objects found by at least one of these three methods with a significance
S/N$ > 4.5$. In order to clean the PSZ2 catalogue from the spurious detections
associated with galactic diffuse emission, some additional restrictions have
been added into the pipeline, together with the removal of those detections
confirmed to be spurious by the PSZ1 follow-ups. More detailed description of
the PSZ2 catalogue construction can be found in \citet{plPSZ2}.

Thanks to the existing surveys and previous intensive follow-up campaigns, about
73\,\% of the sources (1203) in PSZ2 are confirmed clusters, from which 66\,\%
(1094) have redshift estimates. The detailed information about the follow-up
observations and cross-correlation with ancillary catalogues at different
wavelengths is reported in Sec.~7 of \citet{plPSZ2}.

At the time of the catalogue publication, among unconfirmed 559 sources (both PSZ1 and PSZ2), 
450 objects had no validation from other wavelengths.
From them 350 cluster candidates were unique for PSZ2 catalogue 
(i.e. not detected previously in PSZ1 dataset). 

\begin{figure}[h!]
\centering
\includegraphics[width=\columnwidth]{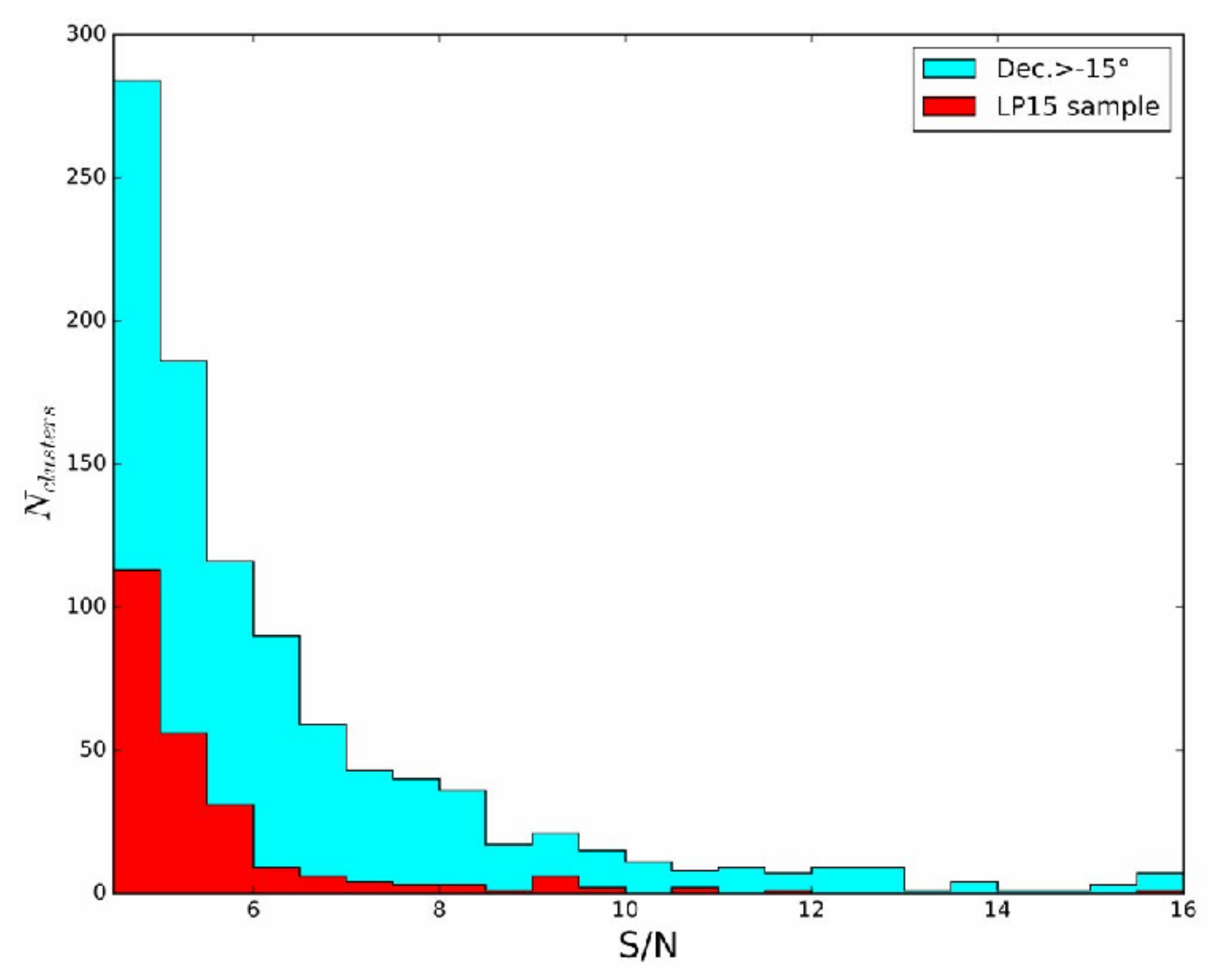}
\caption{Distribution of PSZ2 objects as a function of the S/N of the SZ
detection, and sample definition for this program. For displaying purposes, we
only show the S/N range up to 16, as this is the maximum S/N value in our
sub-sample. The complete sample of sources at Dec.$>-15^{\circ}$ (represented
in light blue) has additional 26 clusters uniformly distributed in the range
$16<$S/N$<50$. The sources observed during {\tt LP15} are shown in red. The
bin size is 0.5.}
\label{fig:snr}
\end{figure}

\subsection{Sample definition and observational strategy}
\label{sec:motiv}

The main motivation of our observational effort was to carry out a systematic
follow-up of the complete set of PSZ2 cluster candidates in the northern sky,
with no confirmed counterparts at the moment of the catalogue publication. The
validation information available at that time was included in one of the columns
of the PSZ2 catalogue, named {\tt validation}, and contained a summary of all
the external identifications. For the definition of the LP15 sample, we choose
all sources in PSZ2 with {\tt validation = -1} (i.e. no known external
counterpart), {\tt PSZ1 = -1} (i.e. no matching detection in the PSZ1
catalogue), and declination above $-15^{\circ}$ (to be easily accesible from the
Canary Islands Observatories). This corresponded to 190 targets (out of the 350
unvalidated all-sky sources).

The reason why we excluded the PSZ1 targets from this sample (48 in total) is
that all those objects are already part of a dedicated follow-up campaing
\citep{barrena18}, and thus, they have been already observed by our team.

This set of LP15 sources was observed during four semesters, 2015B--2017A,
within the frame of the long-term programme {\tt 128-MULTIPLE-16/15B} (hereafter
{\tt LP15}). In total,  106 targets were studied during the first year of the
programme (semesters 15B and 16A), while the remaining 83 were observed during the
second year (semesters 16B and 17A).

All the observations were carried out at the Roque de los Muchachos
Observatory (ORM) on the island of La Palma (Spain) using the following
telescopes: the 2.5\,m Isaac Newton Telescope (INT), the 3.5\,m Italian
Telescopio Nazionale Galileo (TNG), and the 10.4 m Gran Telescopio Canarias
(GTC).  For these three telescopes, we requested in total 44 nights,
approximately 50\,\% of them were dedicated to obtain photometric data (INT) and
other 50\,\% of nights to spectroscopic observations (TNG and GTC). We
summarize main information about {\tt LP15} in Table~\ref{tab:telescopes}. Our
sample of 190 sources corresponded to $\sim 54$\,\% of all unidentified PSZ2
objects, making our programme the largest optical validation campaign of
unconfirmed PSZ2 clusters to date.

Figure~\ref{fig:snr} presents the distribution of cluster counts as a function
of S/N for the full sample of sources observed during {\tt LP15} campaign, in
comparison with the total set of 1003 PSZ2 sources located at
Dec.$>-15^{\circ}$. As it was expected, most of our sources have $S/N <6$, as the
most difficult to validate using some serendipitous shallow surveys available
before our program.

We adopted an observational strategy very similar to the {\tt ITP13-08}
Programme (PSZ1 sources). Before including PSZ2 sources for photometric
observations, we did the initial pre-screening of the proposed targets by
searching for possible counterparts in the Sloan Digital Sky Survey
(SDSS)\footnote{\url{http://skyserver.sdss.org}} DR12 photometric and
spectroscopic data \citep{streb18}. If a cluster counterpart was already
confirmed in the SDSS data, new imaging observations were not required in our
{\tt LP15} programme, and the cluster was directly considered for spectroscopy
with the aim of obtaining its mean redshift, velocity dispersion and dynamical
mass. Galaxy cluster members with SDSS spectroscopic information were also
considered for the mean cluster redshift calculation. After such pre-screening,
we included all the unidentified PSZ2 sources as targets for deep imaging using
$g^\prime$, $r^\prime$ and $i^\prime$ broad-band filters.

In particular, for the subset of 106 sources studied during the first year of
the LP15 programme and included in this paper, 14 of them were already discussed
in \citet{streb18}. Of those, 9 had already a preliminary redshift estimation,
and 5 were considered potential associations.

If cluster counterparts were identified using either our images or the SDSS
data, we performed spectroscopic observations using multi-object spectroscopy
(MOS). We used the GTC telescope to observe the most distant cluster candidates
(at $z_{\rm phot} > 0.35$), while the TNG was used for the nearest ones (at
$z_{\rm phot} < 0.35$).  As the last step, taking into account all the photometric
and spectroscopic information, the cluster validation was performed using the
selection criteria outlined in Sect.~\ref{sec:new}.

\begin{figure*}[ht!]
\centering
\includegraphics[width=\columnwidth]{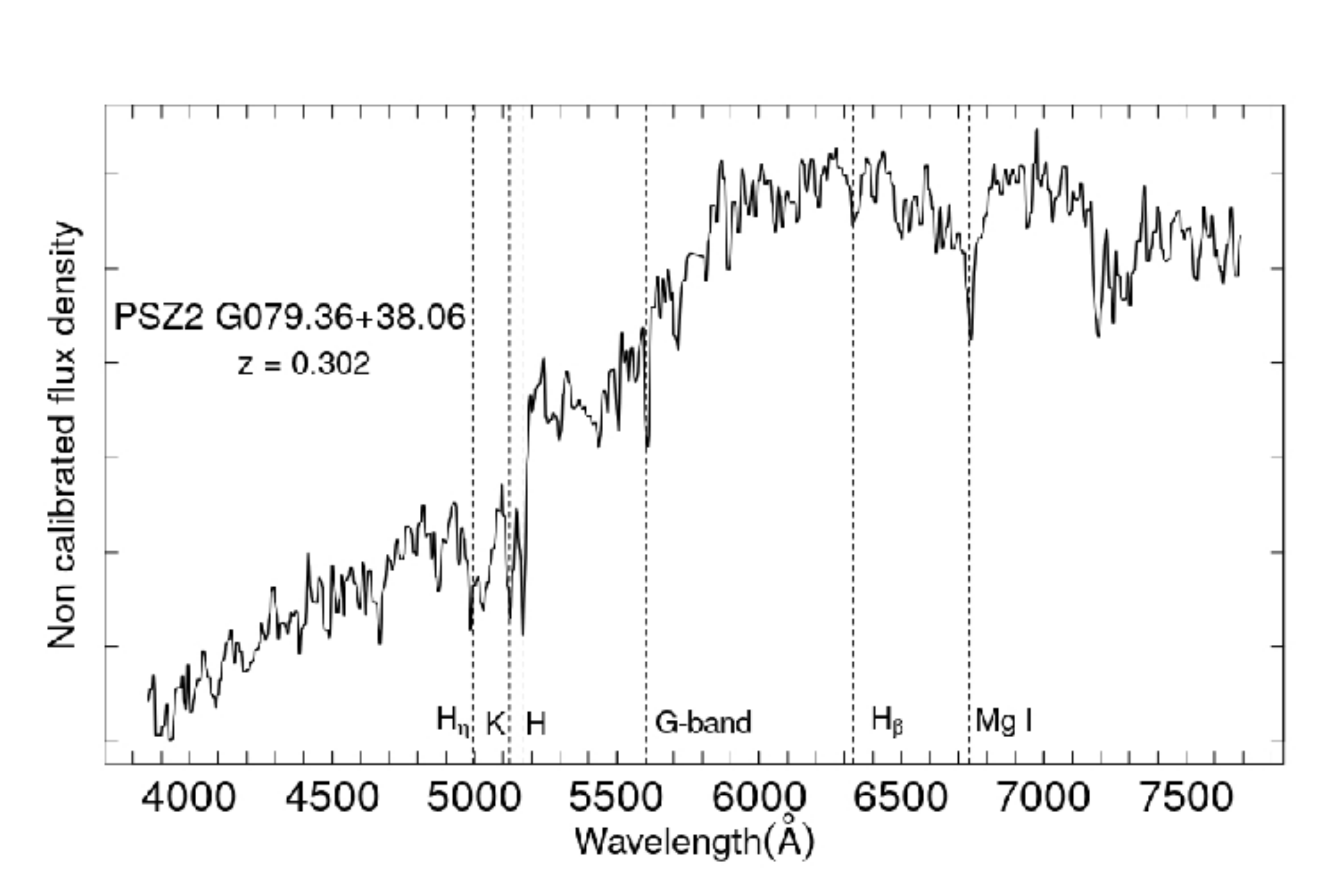}
\includegraphics[width=\columnwidth]{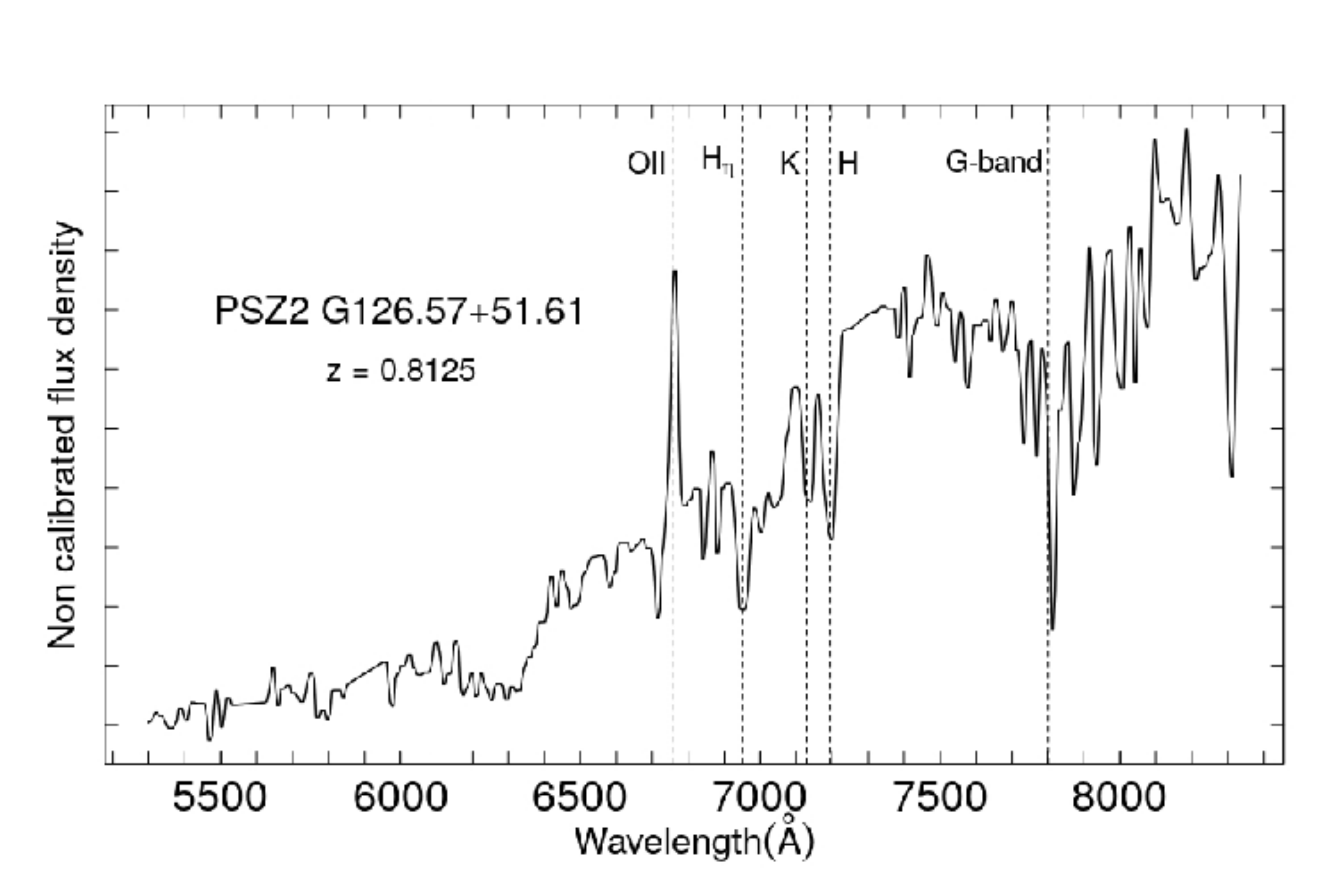}
\caption{Example of the spectra obtained with TNG/DOLORES (left panel) and
GTC/OSIRIS (right panel) for two galaxy members, with magnitudes $r^\prime$ =
18.7 and 21.8, in the clusters PSZ2 G079.36+38.06 ($z = 0.299$) and PSZ2
G126.57+51.61 ($z = 0.816$), respectively. Dashed lines correspond to the
wavelength of the absorption features identified in each spectrum at the
redshift of the clusters. Flux density is plotted in arbitrary units. }
\label{fig:spec}
\end{figure*}

\subsection{Imaging observations and data reduction}
\label{sec:imagedata}

Imaging observations were obtained using the Wide-Field Camera (WFC) installed
in the 2.5\,m Isaac Newton Telescope (INT). The WFC camera at the INT is a four
CCDs mosaic with a field of view (FOV) of $34\arcmin \times 34\arcmin$ and a
pixel scale of $0.\arcsec33$.  To acquire the images, we performed a small
dithering technique of three points with offsets of $10\arcsec$, in order to
clear the resultant image from bad pixels, vignetting and fringing effects and
be able to minimize the impact of cosmic rays. The exposure times vary between
900\,s and 1500\,s per band depending on the magnitudes of the galaxies observed,
which yields to completeness and limit magnitudes  (90\% fractional completeness) in $r^\prime$-band of 22.2 and
23.6 respectively. The seeing conditions also vary from $0.\arcmin8$ to
$1.\arcmin8$.  The photometric data were reduced using standard {\tt
IRAF}\footnote{{\tt IRAF} (\url{http://iraf.noao.edu/)} is distributed by the
National Optical Astronomy Observatories, which are operated by the
Association of Universities for Research in Astronomy, Inc., under cooperative
agreement with the National Science Foundation.} routines. 
The astrometric solution was implemented using the {\tt images.imcoords} IRAF tasks and the USNO B1.0 catalogue \citep{Monet03} as a reference. 
The final astrometric surface provided a precision of about 0$.\arcsec$02 in the centre of the fields, and only in the very 
outskirts the astrometric error  was about 1$\arcsec$. Therefore, the mean rms obtained across the full field of view was  $\sim$ 0$.\arcsec$2.

The photometric calibration refers to SDSS photometry and SDSS standard
fields. Images obtained during non-photometric nights were calibrated in
posteriors runs. In order to detect the sources in the images, we have used {\tt
  SExtractor} \citep{bertin1996} in single-image mode. We detected sources in
$g^\prime$-, $r^\prime$- and $i^\prime$-bands with S/N $\sim 3$ in at least 10
adjacent pixels that implies a 1.5$\sigma$ detection thresholds in the filtered
maps. Using the {\tt MAGAUTO} mode, elliptical aperture photometry was performed
setting the {\tt Kron factor} and the {\tt minimum radius} to the default values
(2.5 and 3.5 respectively).  Finally, the resultant catalogues were merged to
create a master catalogue containing the information of every band.

We also used images in $g^\prime$-, $r^\prime$- and $i^\prime$-bands to create
the deep RGB images which have been used for the visual inspection in our
validation work (see Sec.~\ref{sec:new}).

The broad-band images used to carry out this work has been included in the
Virtual Observatory (VO)\footnote{\url{http://www.ivoa.net/}} collection for
public access. In the near future, our photometric and spectroscopic catalogues
will be also available through this facility.

\begin{table} 
\caption[]{Confirmation criteria adopted to validate or reject clusters
as counterparts of SZ detections (see text for details).} 
\label{tab:crit}
\begin{tabular}[h]{cccc}
    \hline \hline
Flag & MOS Spectroscopy & $\sigma_v$ limit (km\,s$^{-1}$) & $\sigma_R$  \\
 \hline \hline
 1&  YES& $>$ 500 for 0$<z<$0.2 & --   \cr
    &     &    $>$ 650  for $z>$0.2 & -- \cr
 \hline   
2&  NO & NA & $>$1.5  \cr
 \hline   
3& YES &  $<$ 500 for 0$<z<$0.2 & --   \cr
    &      &    $<$ 650  for $z>$0.2 & -- \cr 
    &  NO  &  --          & $<$ 1.5 \cr   
 \hline   
ND &  --  & --           & --        \cr
 \hline   
  \end{tabular}
\end{table}

\subsection{Spectroscopic observations and data reduction}
\label{sec:specdata}

Spectroscopic observations were obtained using the multi-object spectrographs
DOLORES (TNG) and OSIRIS (GTC).  DOLORES (Device Optimized for the LOw
RESolution) is a low resolution spectrograph and camera mounted in the Nasmyth B
focus of the TNG. In MOS mode, it can carry up to 5 masks allowing us to include
between 40 and 50 slitlets per mask. It has a CCD of 2048 $\times$ 2048 pixels
with a pixel size of 13.5\,$\mu$m and a plate scale of
0$.\arcsec$252$/pixel$. We obtained the data using the LR-B grism which provides
a resolution of $R = 600$, a dispersion of 2.75\,$\AA/pixel$ and operates
between 3800 and 8500\,$\AA$. We obtained Hg-Ne and He arcs in order to make the
wavelength calibration of the spectra achieving a \textit{rms} error below 0.1
$\AA/pixel$ over the whole wavelength range. We exposed typically $3 \times
1800$\,s per mask but, depending on the magnitude of the galaxies observed, this
time could change.

OSIRIS (Optical System for Imaging and low-Intermediate-Resolution Integrated
Spectroscopy) is a low and intermediate resolution spectrograph and camera
located in the Nasmyth-B focus of the GTC. Its MOS mode allows up to 70 slitlets
per mask with a typical length of 4--5$\arcsec$. The instrument is composed of a
double CCD of $2048 \times 4096$ pixels with a pixel size of 15\,$\mu$m and a
plate scale of 0$.\arcsec$13$/pixel$. In this work we have used the R300B grism
which operates in the range 4000--9000\,$\AA$ and gives a dispersion of
5.2\,$\AA/pixel$ using the $2 \times 2$ binning set-up ($R \sim 500$). We
obtained Hg-Ar, Ne and Xe arcs in order to make the wavelength calibration of
the spectra achieving a \textit{rms} error below 0.2\,$\AA/pixel$ over the whole
wavelength range. We exposed typically 3$\times$1000\,s per mask obtaining
typical S/N$\sim 5$ for galaxies with magnitudes $r^\prime = 21.6$.

We designed the masks by using previous images obtained for each field in the
corresponding instrument. We used RGB images (composed by $g^\prime$-,
$r^\prime$- and $i^\prime$-bands taken in the INT) as a reference and we included
slitlets with galaxies considered cluster likely members, meaning with coherent
colours and laying in the red-sequence of the clusters (see
Sec.~\ref{sec:pho}). By using these criteria, actual cluster members were
selected with a success rate of typically 50--60\,$\%$ in the inner regions of
the cluster while in the outer regions ($> 0.3$\,Mpc from the brightest cluster
galaxy, BCG) the success rate was around 20\,$\%$.

The reduction process of the spectra followed the subsequent steps: combination
of images, subtraction of sky component from each slit, extraction of spectra,
cosmic ray rejection and finally wavelength calibration. Every step was carried
out using standard {\tt IRAF} tasks. We decided not to apply bias and flat-field
corrections because it adds additional noise to the spectra without significant
improvement of S/N, and our final goal is only the redshift determination. We
used Hg-Ne and He-Ne arcs for DOLORES and Hg, Ne and Ag arcs for GTC to do the
wavelength calibration. We searched for possible deviations in the calibration
using the OI telluric line (5577.3\,$\AA$). We found no systematic offsets but
random fluctuations smaller than 1\,$\AA$, which correspond to $\sim
50$--$80$\,km\,s$^{-1}$ depending on the cluster redshift.

Based on the photometric redshift ($z_{\rm phot}$) estimation (see
Sect.~\ref{sec:pho}), we divided our sample into two sub-samples. The nearby
clusters with $z_{\rm phot} < 0.35$ were observed at the TNG whereas the distant
ones with $z_{\rm phot} \geq 0.35$ were observed at the
GTC. Figure~\ref{fig:spec} shows examples of spectra acquired with both
telescopes.

In order to obtain the radial velocities of the galaxies, we used the task {\tt
RVSAO}\footnote{RVSAO was developed at the Smithsonian Astrophysical
Observatory Telescope Data Center.} implemented in {\tt IRAF}. This routine is
based on the cross-correlation technique developed by \cite{Tonry79}. The method
consists in performing a cross-correlation between the spectrum of our galaxies
and six templates \citep{Kennicutt92} of different galaxy type: E, S0, Sa, Sb,
Sc and Irr.  For each galaxy we adopt the radial velocity that corresponds to
the higher value of the parameter {\textit R} that measures the S/N ratio of the
cross-correlation peak. In general, this method worked properly but in some
cases the procedure led to a non-realistic $z_{\rm spec}$, mainly due to the
low S/N of the spectra.  For this reason, we inspected by eye every result and
checked that it was done accurately. In the majority of the spectra, absorption
lines were present (mainly H and K CaI doublet, H$_{\beta}$, G-Band and MgI
triplet) when they fall on the wavelength range depending on each galaxy
redshift. In a few cases, we could distinguish emission lines such as OII and
OIII doublet which were used to determine the redshift.

The cross-correlation technique together with the quality and the spectral
resolution of the spectra yields a mean error in the radial velocity estimation
of $\Delta v \sim 75$\,km\,s$^{-1}$. However, taking into account double
redshift estimations for a set of around 50 galaxies we were able to determine
the systematic errors. Making a comparison of the two velocity estimates we
obtained a \textit{rms} of $\Delta v \sim 110$\,km\,s$^{-1}$.

The benefits of using multi-object rather than long-slit spectroscopy is the
fact that not only we can determine more precisely the mean redshift of the
clusters but also we can infer their velocity dispersion. On average, we
retrieved between 10 and 25 cluster members per mask. In a first approximation,
a galaxy is considered to be a member of the cluster if its radial velocity lays
within 2500\,km\,s$^{-1}$ in rest frame from the cluster mean radial velocity.
Then, we follow an iterative method considering galaxies as members if their
radial velocity is less than 2.5 times the velocity dispersion away from the
cluster mean velocity. In this way we guarantee that the majority of the
selected galaxies  are members minimizing the presence of interlopers. We note
that, given the FOV of the different facilities, we are usually selecting
members for high-z clusters within the region $R<R_{200}$, where $R_{200}$ is
the virial radius defined as the radius enclosing over-density of 200 with
respect to the critical density of the universe.

\begin{figure}[ht!]
\centering
\includegraphics[width=\columnwidth]{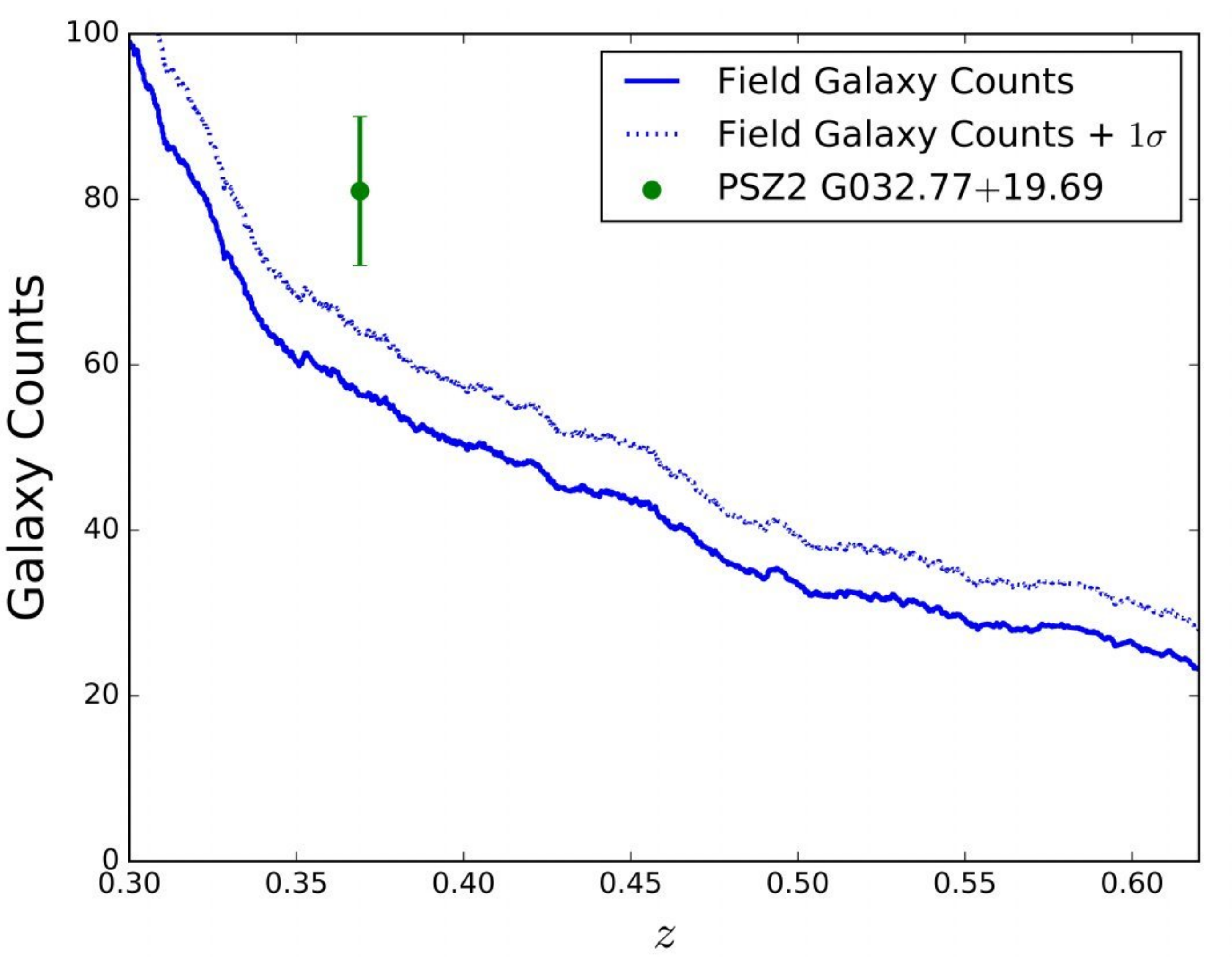}
\caption{Illustration of our methodology to compute the cluster richness using PSZ2 G032.77+19.69 cluster detected at  $z = 0.369$. 
The panel presents galaxy counts as a function of
redshift for this observed field.  The blue line represents the galaxy counts
outside 1\,Mpc region from the optical center of the cluster and the dashed
blue line represents 1-$\sigma$ above the latter.  The green point shows the
galaxy counts (or $R_{0}$, i.e., our initial value of the richness) for this
particular cluster and its 1-$\sigma$ error bars. The corrected value of the
richness $R_{\rm cor}$ is then calculated by subtracting from $R_{0}$ the
background galaxy counts detected at the redshift of the cluster. The complete
description of the calculations and discussion of these clusters are presented
in Sec.~\ref{sec:pho}.  }
\label{fig:figrich}
\end{figure}

\section{Methodology for the identification and confirmation of PSZ2 clusters}
\label{sec:new}
We identified, validated and characterized galaxy clusters using the same
criteria already adopted in our previous papers \citep{planck2015-XXXVI,
barrena18, streb18}. In the next subsection, we will describe again those
criteria, when applied to our {\tt LP15} sample. In short, for each cluster we
use the available photometric data, we carry out the visual inspection, the
inspection of the Compton $y$-map\footnote{MILCA full mission Compton $y$-map
can be downloaded from the {\tt Foreground maps/Compton-SZmap} section
located at \url{http://pla.esac.int/pla/#maps}} \citep{planck2015_xxii}, and
the analysis of the cluster red sequence (using our photometric redshift
estimation). The detected over-densities are evaluated and classified using a
richness parameter. Those clusters with spectroscopic data were also evaluated
using the value of the calculated velocity dispersion of cluster members.

\begin{figure}[ht!]
\centering
\includegraphics[width=\columnwidth]{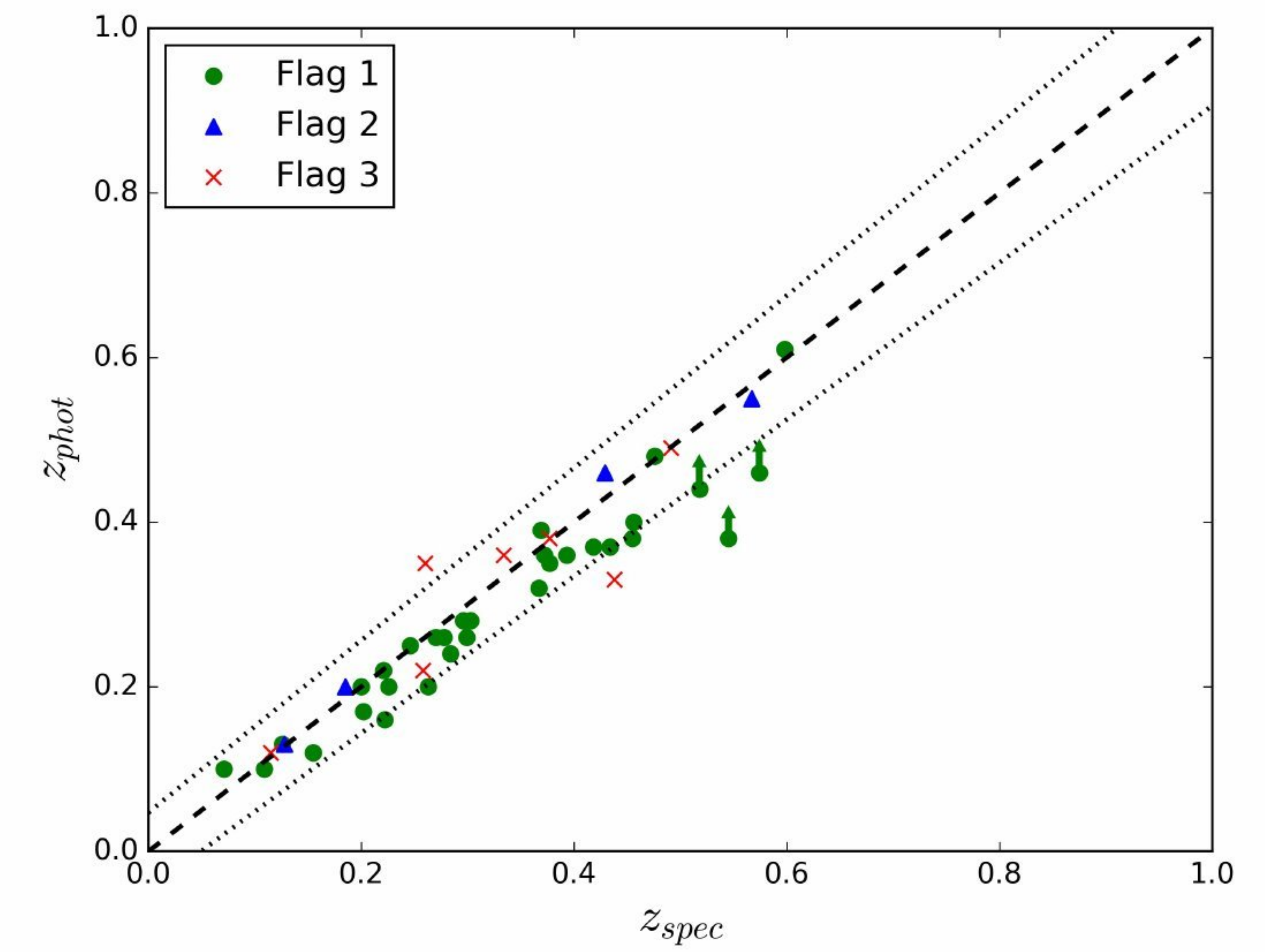}
\caption{Comparison between derived photometric and spectroscopic redshifts for
a sample of sources validated with {\tt Flag}= 1--3.  The dashed line shows
the 1:1 relation. The dotted lines represent the photometric error $\delta_{z}
/(1+z) \sim 0.047$. We underestimate the photometric redshift for clusters at $z >$ 0.5  due to lack of $z$-band photometry in our follow-up. Therefore we assume these values as lower limits. }
\label{fig:zphot}
\end{figure}

\subsection{Photometric validation criteria}
\label{sec:pho}

Our validation steps provide quantitative criteria for a robust association
between PSZ2 source and the observed optical cluster. The visual inspection of
deep RGB images around the official \Planck\ position allows direct
identification of clusters and rich groups in the redshift range $0.1 < z < 0.8$
as a concentration of galaxies of the same colour. In addition, we always
inspect the flux density contours observed in the Compton $y$-map
\citep{planck2015_xxii}, and compare them with the positions of the possible
optical association.  These maps, constructed from linear combinations of the
individual \Planck\ frequency charts, preserve the SZ signal and cancel the
influence of the CMB and galactic emission.  For most of the clusters we observe
the direct dependence between the peak of SZ signal, shifted sometimes from the
\Planck\ PSZ2 nominal source position, and the optical counterpart
\citep{streb18}.  Also, if the location of detected over-density is above the
expected uncertainty in the \Planck\ detection \citep[$\sim 5\arcmin$, see
  Fig.~3 in][]{planck2015-XXXVI} then the structures observed in the $y$-map
allows us to confirm or reject the association.

Once the clusters were identified, we inspected colour-magnitude diagrams
looking for the cluster red sequence (RS) \citep{2000AJ....120.2148G}, using the
colour-magnitude diagrams ($g^\prime - r^\prime$,$r^\prime$) and ($r^\prime -
i^\prime$,$i^\prime$). We fit the RS considering all galaxies with colours
within the range $\pm 0.05$ with respect to the colour of the BCG. We derive the
photometric redshift of the galaxy over-densities following equations (1) and (2)
from \citet{planck2015-XXXVI}. After estimating the $z_{\rm phot}$, we evaluated
the richness of detected systems to validate the observed galaxy clusters as
actual SZ counterparts.

\begin{figure*}[ht!]
\centering
\includegraphics[width=\columnwidth]{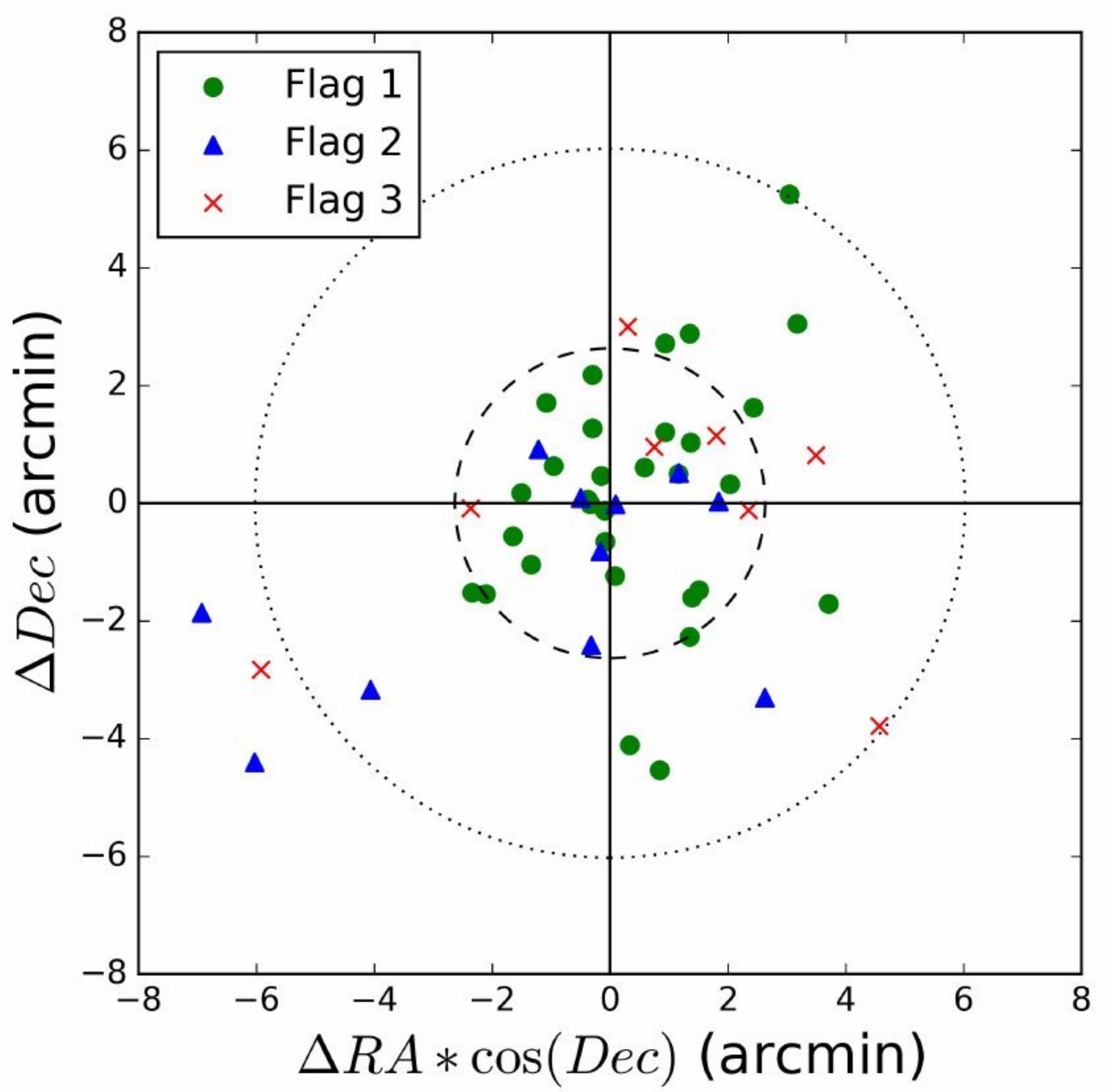}
\includegraphics[width=\columnwidth]{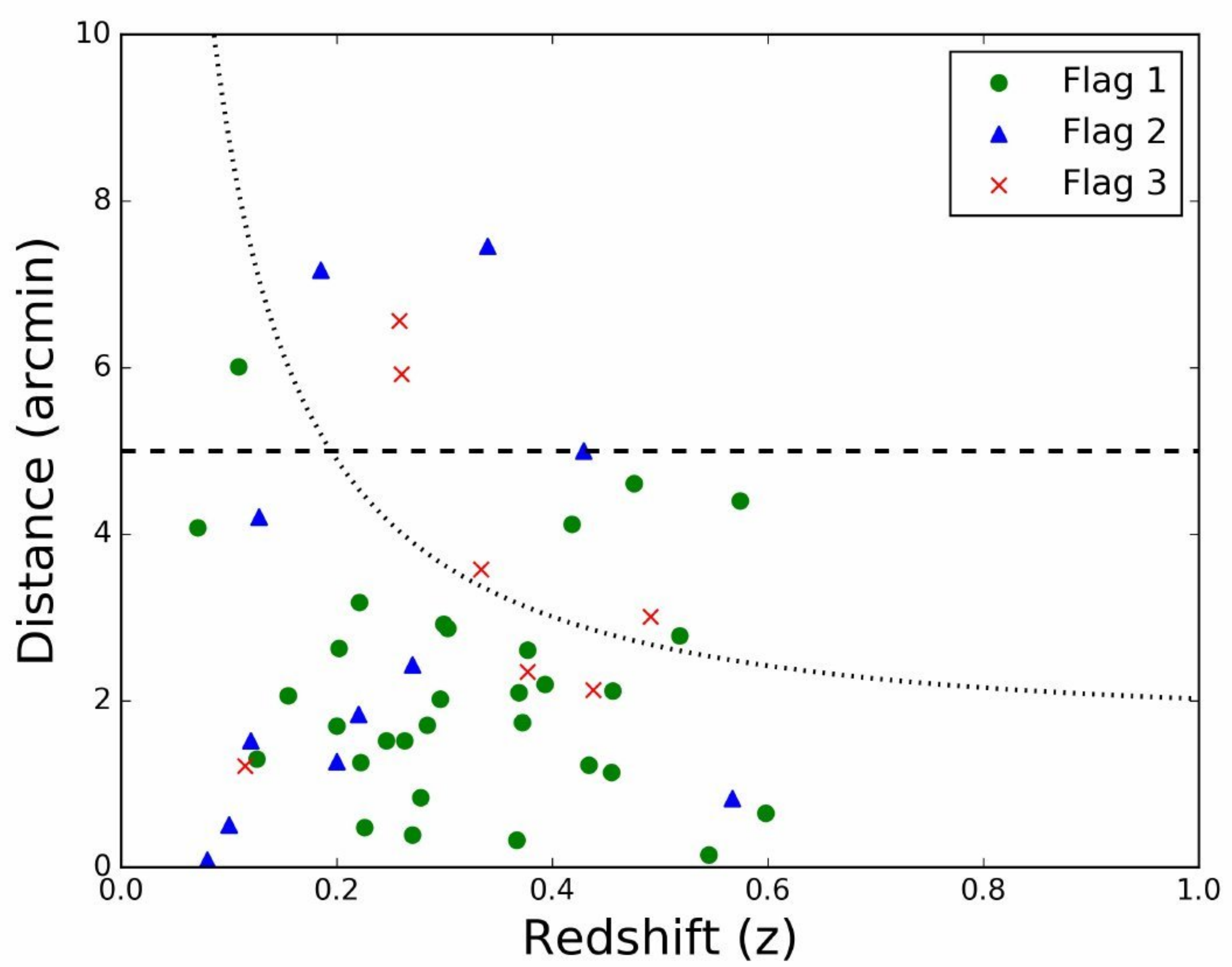}
\caption{Left: Distribution of the optical centre offsets relative to their
\Planck\ SZ positions for the validated clusters ({\tt Flag}= 1--3) presented
in Table \ref{tab:newpsz2}. The inner dashed line corresponds to 2$\arcmin$.6
radius region, which encloses the 68\% of the PSZ2 confirmed
clusters. External dotted line encloses the 95\% of clusters and corresponds
to 6$\arcmin$. Cases with multiple optical counterparts have been excluded
from this analysis.  Right: Cluster optical centre offsets relative to their
\Planck\ SZ position as a function of cluster redshift for a sample of 50
sources. The dashed horizontal line is 5$\arcmin$, which represents the
maximum offset expected for a \Planck\ SZ detection (i.e. FWHM
\Planck\ multi-frequency combined beam). The dotted line corresponds to the
physical 1\,Mpc radius region at the corresponding redshift.  }
\label{fig:sdss_z3}
\end{figure*}

The PSZ2 survey selection function \citep[see Fig.~26 of][]{plPSZ2} shows that
the expected \Planck\ SZ detections are massive systems, with a mean mass over
the whole redshift range of $4.82 \times 10^{14}$\,M$_\odot$. Therefore, we
would expect our sample to be composed by massive rich clusters, and thus, no
poor systems should in principle be validated if they are found along the line
of sight of the \Planck\ detection. In order to adopt an objective criterion for
discarding low mass systems, we defined a richness parameter ($R$).  There are
multiple approaches in the literature to calculate the richness of the observed
systems. In this paper, we present a refined proceduce with respect to our
methodology in previous works \citep{barrena12}, taking into account the local
background variance and making it more robust in comparison with other
validation works in the literature. In short, $R$ is computed as the number of
likely members (galaxies in the RS$\pm 0.15$ magnitude locus) in $g^\prime -
r^\prime$ and $r^\prime - i^\prime$ for clusters at $z<0.35$ and $z \geq 0.35$,
respectively, showing $r^{\prime}-$magnitudes in the range
$[m_{r^{\prime}}^{\star}-1, m_{r^{\prime}}^{\star}+1.5]$, where
$m_{r^{\prime}}^{\star}$ is the characteristic magnitude and depends on the
redshift \citep[see][]{barrena12}. We count galaxies within a projected region
of 1\,Mpc radius from the optical center of the cluster at its redshift. This
initial value of the richness ($R_{0}$) is then corrected for the field galaxy
counts ($R_{\rm f}$), which is computed in the same way but outside the 1\,Mpc
radius region for each cluster. We must stress that this final value ($R_{cor}
\equiv R_{0} - R_{\rm f}$) should be considered as a lower limit to the richness
of the system, as we are formally counting in the $R_{\rm f}$ estimation some
clusters members that might lay outside the 1\,Mpc region.

We based our confirmation criterion in the value $\sigma_{\rm R}$, which is
computed as $R_{\rm cor}/ \sqrt{R_{\rm f}}$, and describes the richness
significance above the local background level. Given that in our calculations
the richness is not a fixed value, and depends on the local environment, we can
in principle equally good validate clusters observed either in crowded star
fields or in empty areas. 

Despite the flexibility and robustness of this approach to calculate the
richness, we observed in two situations the weakness of this method. First, the
method is not working properly when the FOV is relatively small compared to the
cluster size.  Also, it could produce wrong results if we observe an over-density of sources
in the background. The algorithm then calculates artificially a high background $R_{\rm f}$ and, consequently,
shows an underestimated value of $R_{\rm cor}$ due to the over-subtraction of this local background from the data. 
If this is the case, after the careful
inspection of the images, we decided to keep the original (uncorrected) $R$
value (see notes in Table~\ref{tab:newpsz2}).

Fig.~\ref{fig:figrich} illustrates our method for the determination of the
richness using one of the confirmed clusters from our sample. 
The initial values of richness for the
the cluster was $R_{0} = 81$. PSZ2 G032.77+19.69 is located in a crowded star
area, so the richness of the field at the redshift of the cluster is also high
($R_{\rm f} = 56.4$) yielding $R_{\rm cor} = 24.6$ and $\sigma_{\rm R} =
3.28$.

\subsection{Spectroscopic validation criteria}

In addition to the photometric data, we obtained for almost all clusters
spectroscopic information.  This includes our own observation and publicly
available data from the SDSS survey.

We find a good agreement between the photometric and spectroscopic redshifts for
all our PSZ2 sources, except for high-z clusters (see Fig.~\ref{fig:zphot}). We recall
that our photometric redshift is based in the $r^\prime - i^\prime$ colour of
likely cluster members \citep{planck2015-XXXVI}, which is not appropriate
estimator for systems at $z > 0.7$. Obtaining secure redshift
for sources at $0.5 < z_{\rm phot} < 0.7$ is also limited by the lack of $z$-band 
photometry and obtained values must be considered as a lower limit. Our study yields a photometric redshift error of
$\delta_z /(1+z) \sim 0.047$ when considering clusters with $z < 0.7$ ( $\sim 0.04$ for  $z < 0.5$). 

On average, we typically obtain about 20--40 spectroscopic members per cluster,
and, consequently, a velocity dispersion $\sigma_v$ can be estimated. We used
this value of $\sigma_v$ to investigate whether these clusters are poor or
massive systems. There is a direct dependence between the redshift and mass of
the \Planck\ clusters, reported first in \citet{planck15} and studied in detail
for high-z ($z > 0.5$) clusters in \citet{burg16}.

We expect that clusters at $z < 0.2$ with $M_{500} > 10^{14}$\,$M_\odot
\ h_{70}^{-1}$ will present $\sigma_v > 500$\,km\,s$^{-1}$, whereas clusters at
$z > 0.2$ with M$_{500} > 2 \times 10^{14}$M$_\odot \ h_{70}^{-1}$ should show
$\sigma_v > 650$\,km\,s$^{-1}$ \citep{mun13}. As in \cite{barrena18}, we assume
these values in the velocity dispersion to distinguish between actual and
detectable systems by \Planck\ and chance identifications not linked to the SZ
effect.

\subsection{Summary of our validation criteria}

Using all available photometric and spectroscopic information, we adopted a set
of flags according to the validation level of each target (see Table
\ref{tab:crit}). For the sources with the spectroscopic information, we based
our confirmation criteria only on the dynamical properties of the clusters
(through their velocity dispersion). If no spectroscopic information available
or only a few spectra is present, we validate the cluster using the value of
$\sigma_{\rm R}$ for the richness significance above the background.  We noticed
that most of our spectroscopically confirmed clusters have $\sigma_{\rm R}$
larger than 1.5, so we adopt this value as a threshold in our study.

The distance from the \Planck\ nominal pointing is also taken into account,
however we consider this condition as a flexible restriction, especially in the
case of double detections or sky regions contaminated by the galactic dust. The
maximum offset expected for a \Planck\ SZ detection is about 5$\arcmin$ (i.e.,
roughly the angular resolution of the high-frequency HFI channels). In the case
of finding a distance larger than $5\arcmin$ and obvious cluster presence, we
always inspect the distribution of the SZ flux in the Compton $y$-map, and
confirm/reject cluster according to the observed SZ contours. The largest
distance from the Planck pointing, as confirmed by contours from the $y$-map,
was found to be 7.4$\arcmin$ for a cluster at $z_{\rm phot} = 0.34$ (see
Section~\ref{sec:disc} and Fig.~\ref{fig:psz2_583_fin}).

\begin{figure}[ht!]
\includegraphics[width=\columnwidth]{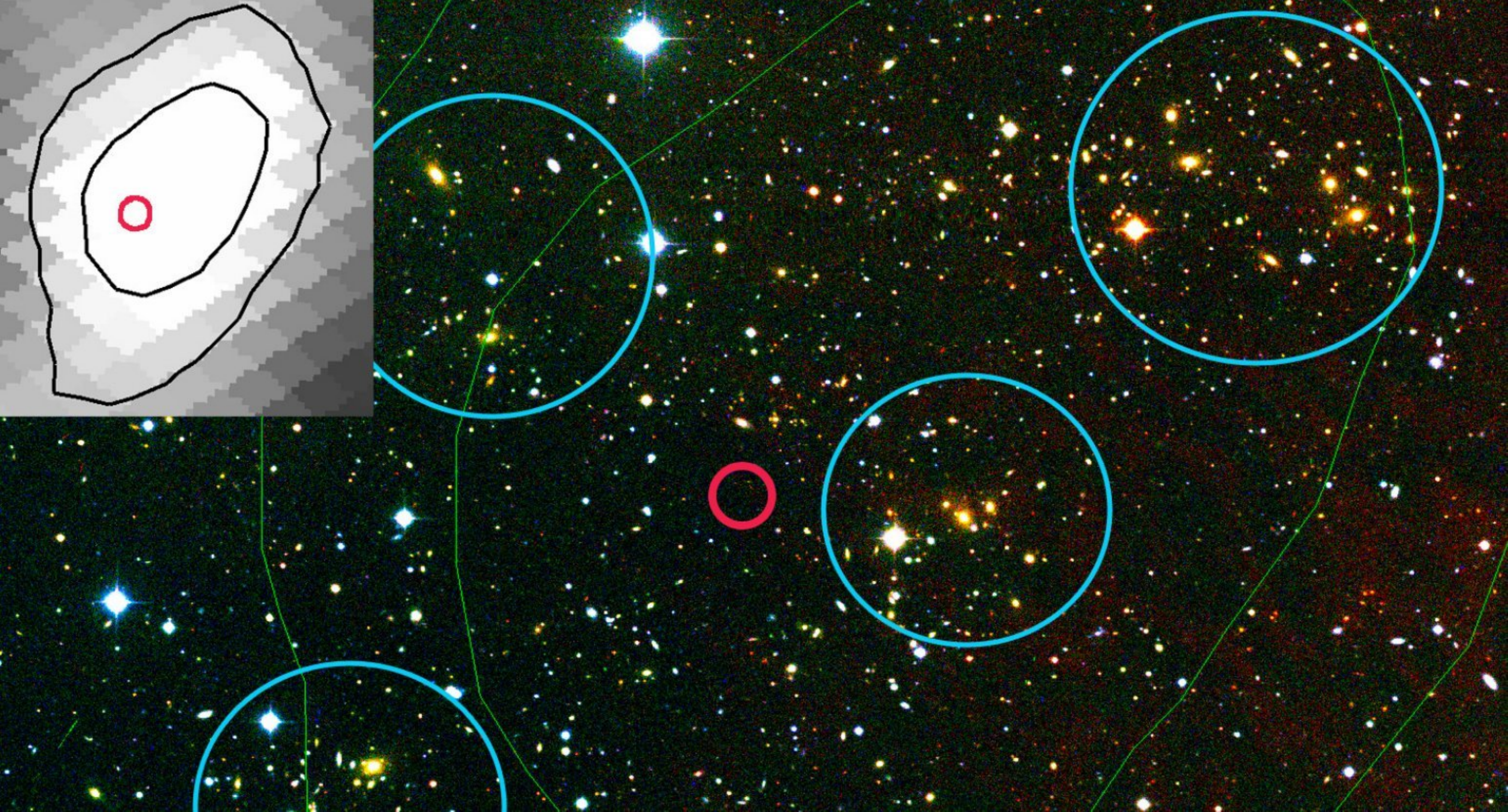}
\caption{The zoomed RGB image of the PSZ2 G079.36+38.06. This cluster is one of
the richest in our dataset, with more than 300 photometric members, 53 of them
confirmed through the spectroscopy at $z_{\rm spec}=0.299$. The observed
galaxies are grouped in several clumps (marked as blue circles) and
distributed across FOV ($\sim 11\arcmin \times 19\arcmin$) of the optical
image.  The top-left panel shows the MILCA $y$-map with black contours
corresponding to the 3 and 6$\times 10^{-6}$ levels of the $y$-map in this
area (in the RGB image the same contours are represented as green lines). The
red circles in both images correspond to the nominal \Planck\ position.  }
\label{fig:psz2_336_fin}
\end{figure}

\begin{figure*}[ht!]
\centering
\includegraphics[width=\columnwidth]{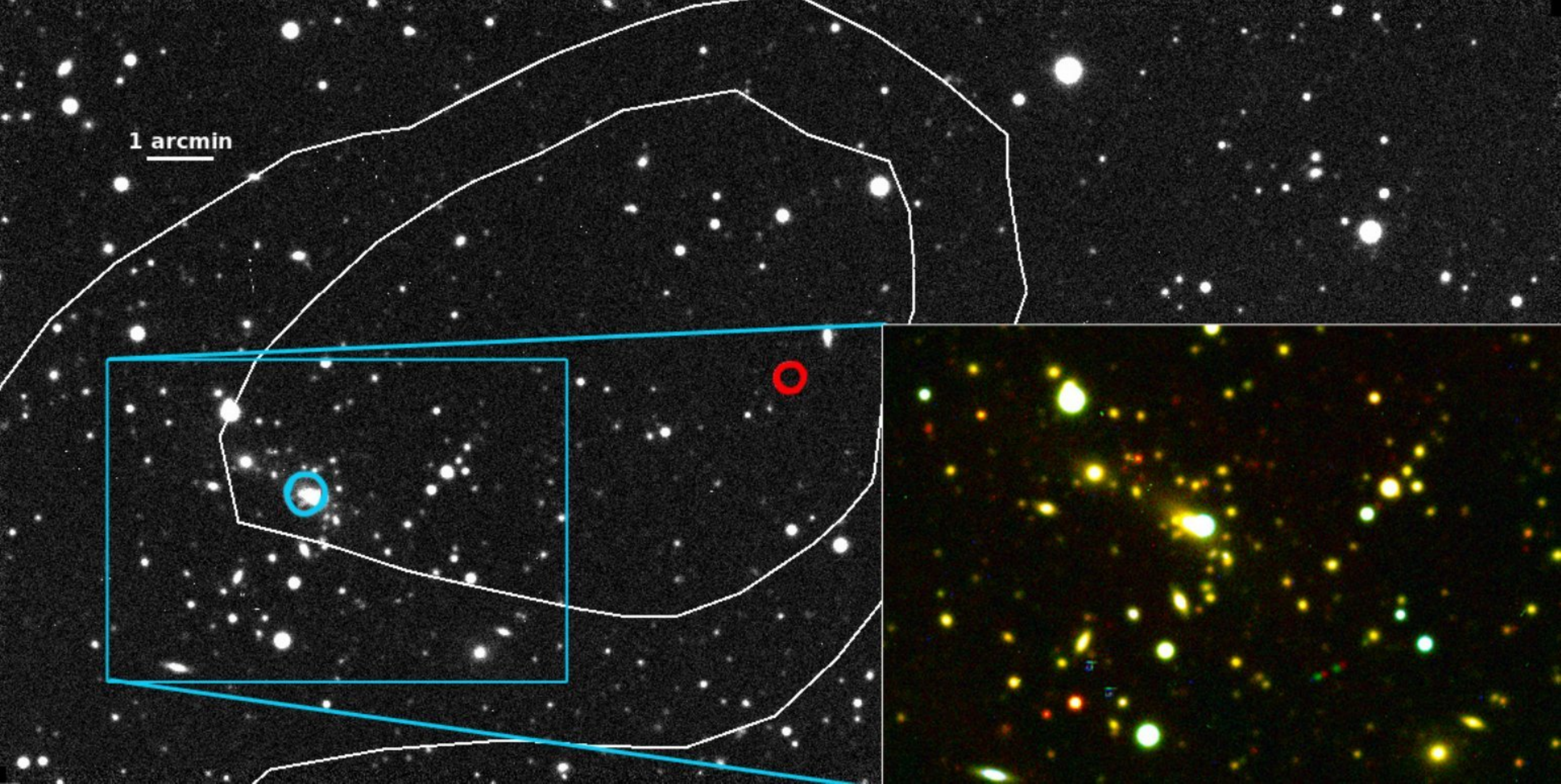}
\includegraphics[width=\columnwidth]{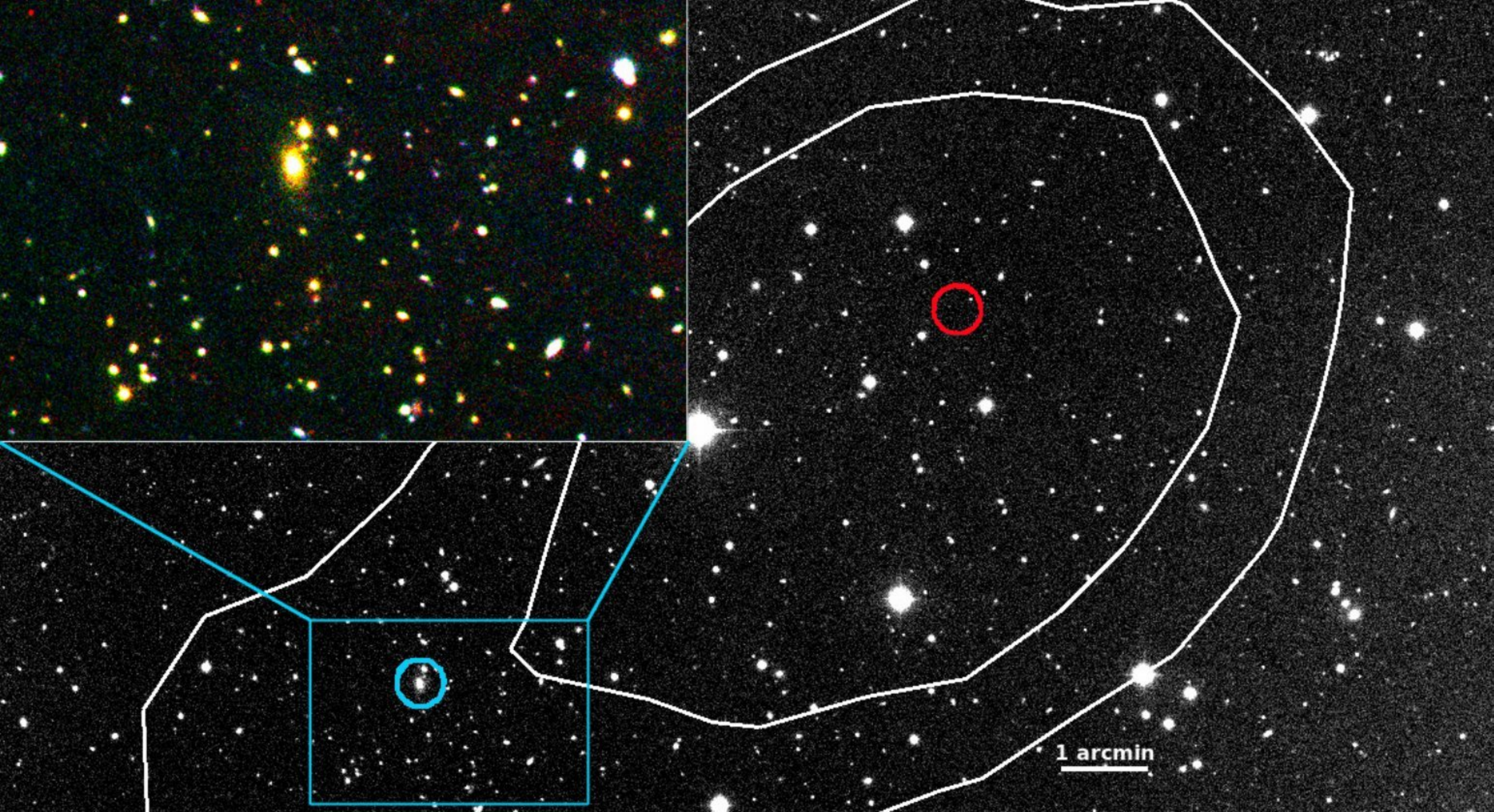}
\caption{Two examples of the clusters where the distance between the optical
counterparts and the nominal PSZ2 position are beyond the official 5$\arcmin$
limit. Left: PSZ2 G084.69-58.60 with BCG at 7$\arcmin$.1 at $z_{\rm spec}=0.185$. 
Right: PSZ2 G118.79+47.50 with BCG at 7$\arcmin$.4 at $z_{\rm phot}=0.34$. For 
both clusters we show WFC/INT $r$-band images with white contours corresponding 
to the 3 and 6$\times 10^{-6}$ levels of the Compton $y$-map in this area. 
The red circle indicates the nominal PSZ2 position. The blue circle marks the 
BCG and the zoomed RGB image shows the central area of these rich clusters. 
The elongated shape of $y$-map contours towards the clusters supports these 
associations.}
\label{fig:psz2_583_fin}
\end{figure*}

\section{Results for the PSZ2 catalogue}
\label{sec:disc}
\subsection{Confirmed clusters}

Table~\ref{tab:newpsz2} summarizes our results for the  106 PSZ2 sources explored
in the first year of our optical follow-up.  We provide the identification
number in the PSZ2 catalogue, the \Planck\ name, S/N of the SZ detection,  neural network quality flag {\tt
  Q$\_$NEURAL}, optical counterpart coordinates (assumed to be those of the BCG position or, in
the absence of BCG, an approximate geometrical centre of the likely members),
distance between the optical and SZ centres (in arcmin), redshift (photometric
and, if available, spectroscopic), and number of spectroscopically confirmed
galaxy members. We also provide the richness information for each cluster,
quoting the richness value and the value of $\sigma_R$ as described above. The
last two columns provide our cluster classification, following the flagging
scheme described in Section~4.3, and some comments relative to other possible
identifications or noteworthy features.

Some of our clusters were pre-selected from \citet{streb18} (mainly for MOS
observations) and, thus, they have SDSS photometric information available. We
also quote for these clusters the additional name from the catalogue of
\citet{wen12}, if available. In summary, we update the information for 14 clusters, 
being 13 of them members of the LP15 sample.
The remaining one is PSZ2 G310.81+83.91, which was also identified in the PSZ1 catalogue, and thus it is not  included in
the LP15 sample.

At the moment of the publication, several articles
reported confirmation of some clusters from our sample. For example,
\citet{boada18} presented photometric redshifts for eight clusters. For all
matched sources, except for three clusters, we have secure spectroscopic
redshift measurements which are in perfect agreement with the reported
photometric information.  For PSZ2 G106.11+24.11, PSZ2 G107.83-45.45 and PSZ2
125.55+32.72 we have only photometric confirmation. Therefore, and for
completeness, we presented in Table~\ref{tab:newpsz2} our alternative
photometric measurements.

Following the confirmation criteria given above, we find that 50 PSZ2 sources
present clear over-densities around the nominal \Planck\ position. However, after
the inspection of obtained $\sigma_v$, we classified eight clusters as weak
associations with the corresponding SZ source (i.e., {\tt Flag}= 3). Thus, in
total, we were able to confirm  41 new PSZ2 sources, 31 of them classified with
{\tt Flag}=1 (spectroscopic confirmation) and 10 with {\tt Flag}=2 (photometric
confirmation). In three cases, we found multiple optical counterparts along line
of sight.

The detailed description of cluster counterparts with the spectroscopic
confirmation and its corresponding physical properties, such as velocity
dispersions and dynamical masses, will be discussed in detail in a future paper.

For 50 detected clusters (with {\tt Flag} 1--3) we studied the dependence
between the position of optical centre and the nominal Planck SZ
coordinates. The position error predicted for SZ detections in the \Planck\ SZ
maps was about $2\arcmin$ for targets in the PSZ1 sample and it was calculated
that the cluster associated with the SZ effect should be closer than $\sim
5\arcmin$ (e.g., the beam size of the SZ detection) from the SZ PSZ1
source coordinates. However, it was shown during the follow-up campaigns
\citep[e.g.,][]{planck2015-XXXVI, barrena18, streb18} that even if this is true
for most of the sources, some small percentage of true counterparts are located
in the distances of $\sim 6$--$8\arcmin$. In most of the cases it corresponds to
nearby systems at $z < 0.25$ with large apparent radius or fields affected by
the optical structures, such as galactic cirri, which influence the
\Planck\ maps and, consequently, the final position of the detected SZ source.
Fig.~\ref{fig:sdss_z3} (left panel) shows the final offset distribution of
cluster optical centre relative to their \Planck\ SZ position.  As it was
expected, 68\,\% of the 50 confirmed cluster sample are enclosed within
2$\arcmin$.6, while for 95\,\% of the sources this corresponds to 6$\arcmin$.
We also studied dependence between this observed offset and redshift of the
cluster (Fig.~\ref{fig:sdss_z3}, right panel). As it was expected, most of the
clusters are located not only inside the 5$\arcmin$ region, but also inside the
physical 1\,Mpc region at the cluster redshift.  In six cases (one of which
corresponds to multiple optical counterpart detection) we observe the true
counterparts at distances$> 5\arcmin$. The largest distance from the Planck
pointing, confirmed by contours from the $y$-map, was 7.4$\arcmin$ (see
discussion below).

\begin{figure}[ht!]
\centering
\includegraphics[width=\columnwidth]{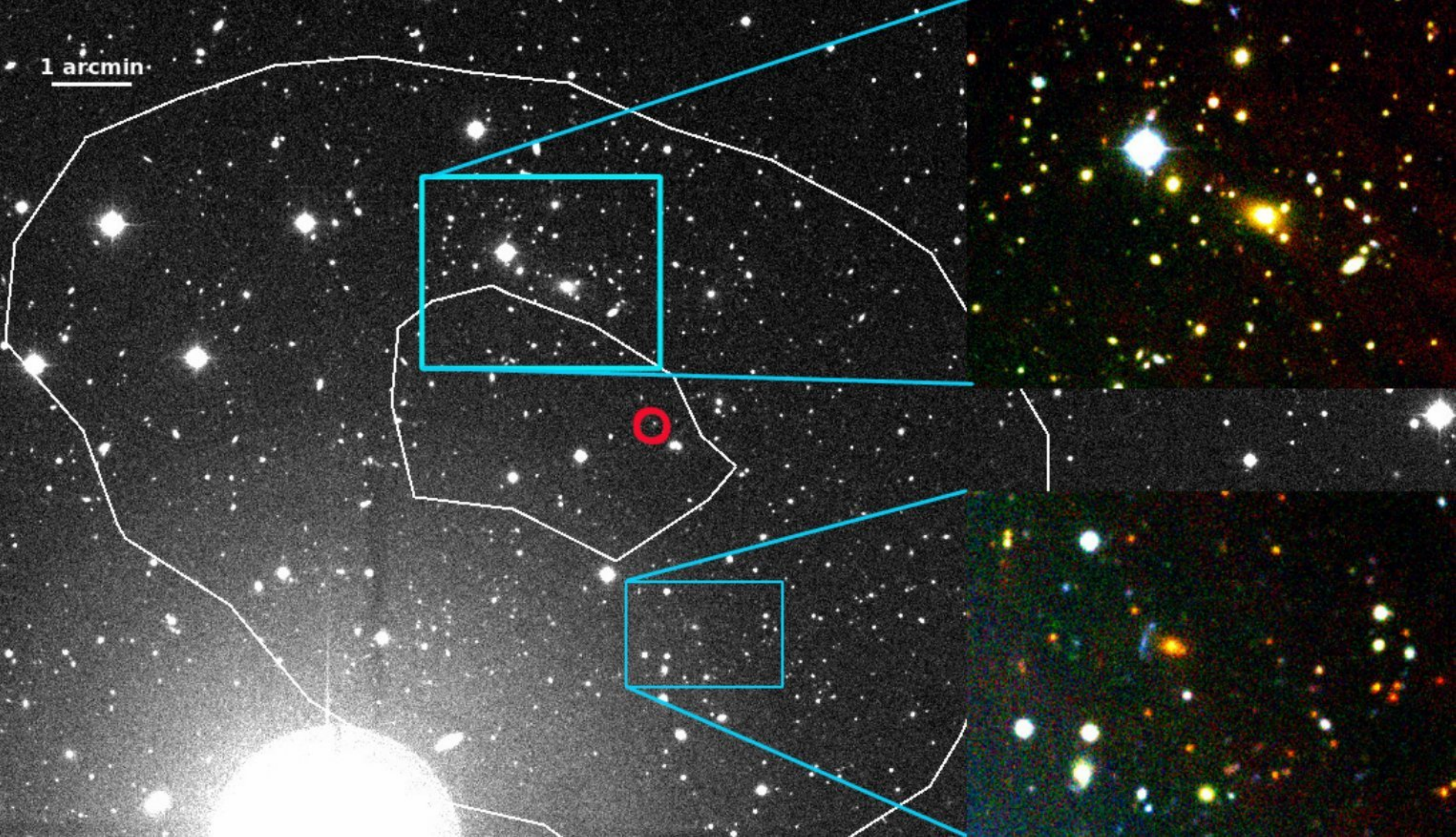}
\caption{WFC/INT r-band image of the PSZ2 G120.76+44.14 which shows the presence
of 2 optical counterparts at different redshifts located at almost the same
distance from the official \Planck\ PSZ2 position ($\sim 2\arcmin$).  The white
contours correspond to the 3 and 6$\times 10^{-6}$ levels of the Compton
$y$-map in this area, and enclosed both observed sources. The small red circle
indicates the position of \Planck\ PSZ2 source.  The small RGB images show the
zoomed regions around the two clusters, one at $z_{\rm spec}= 0.296$ (top
panel), and a second cluster at $z_{\rm spec}= 0.393$ (bottom panel).  We also
detected a gravitational arc around the BCG of the second cluster.  }
\label{fig:psz2_592_fin}
\end{figure}

\begin{figure}[ht!]
\centering
\includegraphics[width=\columnwidth]{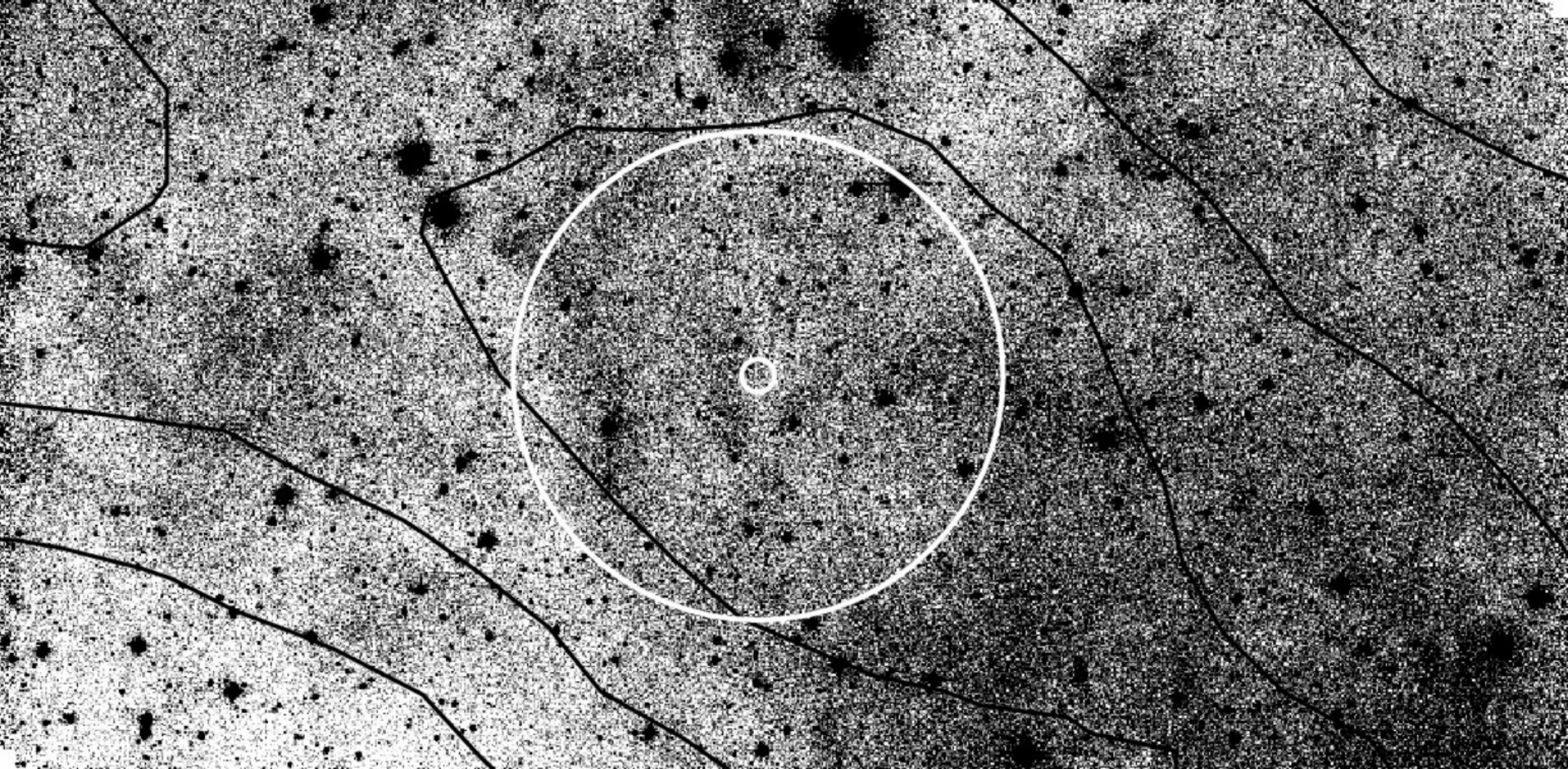}
\caption{WFC/INT g-band image of the PSZ2 G100.07+17.06. The black contours
correspond to the 3, 6, 9$\times 10^{-6}$ levels of the Compton $y$-map in
this area. The small and big white circles indicate the nominal
\Planck\ position and the $5\arcmin$ region, respectively. No cluster
counterpart is identified in this case. Most probably, the presence of
important galactic gas and dust structures influenced on the SZ emission in
this area, creating spurious enhancements of the SZ signal and, consequently,
false SZ detection. We also observe that the elongation and direction of a
dust Galactic filament is strongly correlates with observed SZ flux profile,
supporting our hypothesis. }
\label{fig:non}
\end{figure}

In the following, we describe, as examples, a few clusters showing some
particular features.

\paragraph{PSZ2 G023.87-13.88} is the only fossil\footnote{defined as galaxy
systems with a magnitude difference of at least two magnitudes in the $r$-band 
between the BCG and the second-brightest galaxy within half the virial radius 
$R_{200}$} cluster in our sample. It contains about 25 photometric members at 
$z_{\rm phot}= 0.12$.

\paragraph{PSZ2 G079.36+38.06} This cluster is one of the richest systems in 
our dataset. We detected more than 300 photometric members, distributed across 
the image and grouped in a few clumps (Fig.~\ref{fig:psz2_336_fin}). We were 
able to obtain spectroscopy for 53 sources and confirm $z_{\rm spec}=0.299$ with 
a $\sigma_v=913$\,km\,s$^{-1}$.  An example of one of these spectra is presented in
Fig.~\ref{fig:spec}.

\paragraph{PSZ2 G084.69-58.60 and PSZ2 G118.79+47.50} Even though these 
clusters are beyond the limit of the official 5$\arcmin$ distance accepted for 
SZ sources (7$\arcmin$.1 and 7$\arcmin$.4, for PSZ2 G084.69-58.60 and PSZ2 
G118.79+47.50, respectively) the MILCA contours confirm that these clusters are 
actual counterparts to the SZ signal (Fig.~\ref{fig:psz2_583_fin}).  In total, 
we have six cases where the clusters are located beyond 5$\arcmin$ limit 
(5$\arcmin$.4--7$\arcmin$.4) and in all cases the contours from $y$-maps 
support the validation. For some sources we even observed the shift between 
the peak of the SZ signal ($y$-map) and the nominal \Planck\ position, similar 
to Fig.~8 in \citet{streb18}.

\paragraph{PSZ2 G120.76+44.14} is an example of SZ sources with multiple optical 
counterparts. Both clusters are rich systems and located at a similar distance 
from the nominal \Planck\ position ($\sim$ 2$\arcmin$) and confirmed by $y$-map 
contours. One of these two clusters is a system at $z_{\rm spec}=$ 0.296, while 
the second has $z_{\rm spec}=$ 0.363 and presents a clear gravitational arc around 
the BCG (Fig.~\ref{fig:psz2_592_fin}). Both clusters probably contribute to the SZ
emission.  In total, we find three cases like this, where multiple counterparts
are associated with a single SZ source. We denote these sources with a special
symbol in the Table \ref{tab:newpsz2}.

\begin{figure}[ht!]
\centering
\includegraphics[width=\columnwidth]{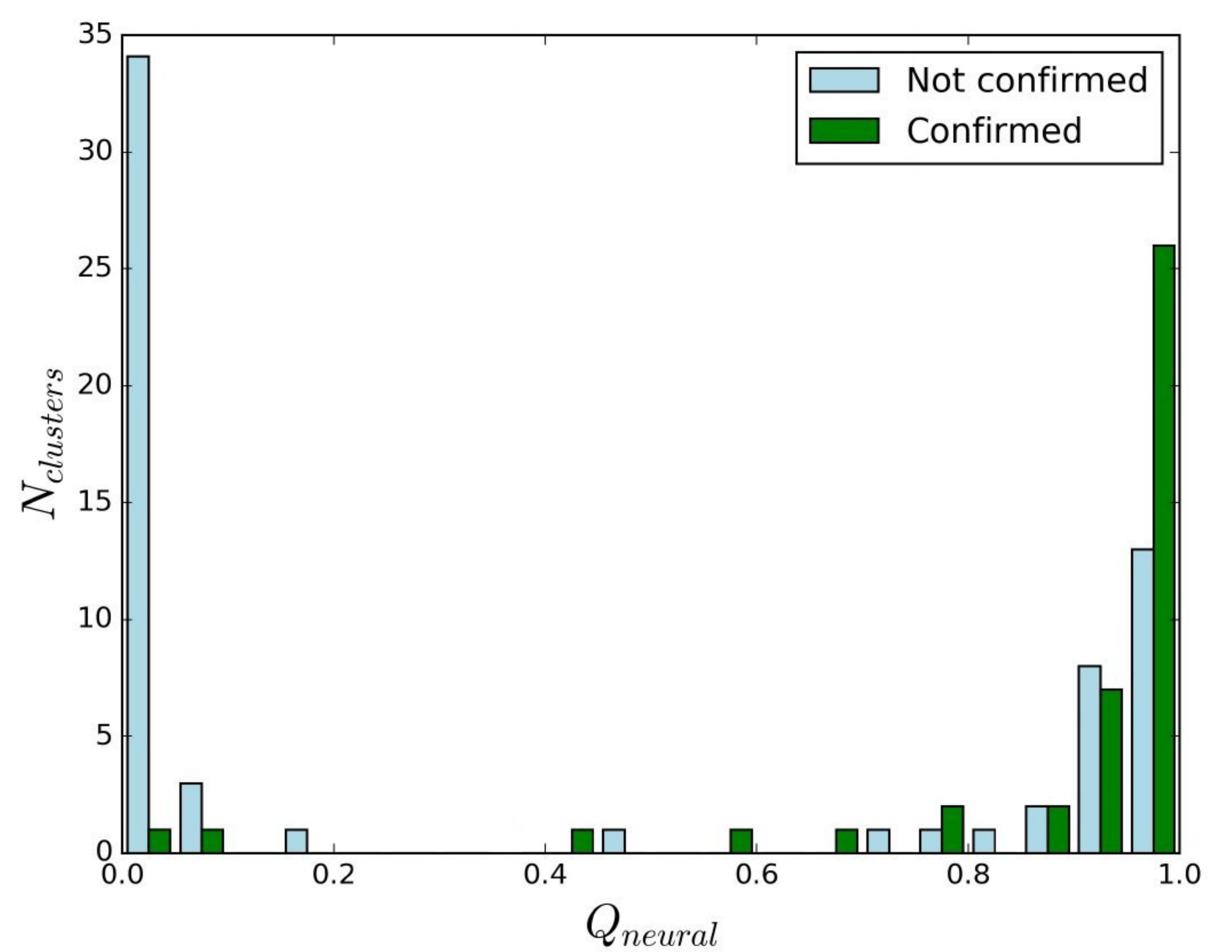}
\caption{Number of cluster-candidates versus the neural network quality flag
value for the subsample of 106 sources studied in this paper. By construction,
{\tt Q$\_$NEURAL} values smaller than 0.4 denote low-reliability
detections. This figure confirms that this parameter indeed effectively
separates real and spurious SZ identifications. However, we note that some of
the unconfirmed sources with high {\tt Q$\_$NEURAL} values are located in
areas of strong dust contamination, probably producing the false SZ
detections. }
\label{fig:qn2}
\end{figure}

\subsection{Unconfirmed cluster candidates}

Out of the 106 sources studied in this paper. 57 remain unconfirmed. Most of
those SZ targets are located close to the galactic plane and are associated with
areas with strong dust contamination, thus, probably, producing false SZ
detections in the \Planck\ maps (see e.g. Fig.~\ref{fig:non}). Similar cases
were already reported and intensively discussed in our previous follow-up works
for PSZ1 targets \citep[Section 4.3 in][]{barrena18} and pre-selected PSZ2
sources \citep[Section 4.3 in][]{streb18}. In the presented field, as in the
majority of other cases, we detect important galactic cirrus around the PSZ2
G100.07+17.06 in the optical images. Moreover, the distribution of the signal in
the $y$-map is not compact, and shows an elongated profile along the observed
Galactic dust structures.

We emphasize here the importance of using optical and infrared observations to confirm the
absence of a cluster counterpart in those regions of strong dust contamination
and for sources detected with relative low S/N. For example, \citet{kha16}
proposed a theoretical approach for the validation of the PSZ2 clusters based on
the combination of CO and $y$-maps only. They classified all sources in groups
based on their ${\chi}^2_{CO-y}$ determining whether a source is a cluster or a
molecular cloud. However, only 61\% of 1094 previously confirmed PSZ2
sources with secure $z$ information were classified as clusters ($CLG + pCLG$)
using this method. The rest of the confirmed clusters were classified either as
indeterminable ($IND$) or as molecular clouds ($MOC + pMOC$). Studying their
classification for our subsample of 106 sources, we found that 55\% of our
confirmed clusters were classified as molecular clouds, and 14\% of our
unconfirmed sources classified as clusters. However, we note that all our high S/N 
sources without an optical counterpart are classified as $MOC$ by this approach.

We note that for some sources, we cannot explain the absence of optical
counterparts in terms of dust contamination. A relatively high number of
non-detections in ``clean'' fields was found already in the PSZ1 sample, and it
is re-confirmed in the PSZ2 set.  Inspecting in detail all our unconfirmed
sources, we found that most of them have neural network quality flag {\tt
  Q$\_$NEURAL} \citep{ag15} close to 0. This flag was introduced already for the
PSZ1 dataset, and the value of 0.4 was used to separate the high quality
detections from the low-reliability ones. The original PSZ2 catalogue included
171 detections considered likely to be spurious by the neural network
clasification, and from them only 19 were confirmed clusters with redshift
information. Our subsample of  106 sources contains  37 objects with {\tt
  Q$\_$NEURAL} $< 0.4$, and all of them, except PSZ2 G137.24+53.93 and PSZ2
G310.81+83.91, were classified as unconfirmed. This result demonstrates again
that this flag effectively separates between real and spurious detections, and
thus it should be considered as valuable information in the validation
process. In Fig.~\ref{fig:qn2} we plot a distribution of confirmed and
unconfirmed sources versus {\tt Q$\_$NEURAL} values. Some of the not validated
clusters with high values of {\tt Q$\_$NEURAL} are located at the areas with
strong dust contamination being, most probably, result of spurious enhancements
of the SZ signal.

Finally, we mention that some of these unconfirmed sources with the low value of
{\tt Q$\_$NEURAL} flag at the same time present a high S/N of the SZ detection
(values greater than 10), like PSZ2 G100.45+16.79, PSZ2 G153.56+36.82 and PSZ2
G107.41-09.57. This again demonstrates the importance to perform
multi-wavelength follow-ups which include optical data, which allow us to
correctly validate SZ sources and establish the actual completeness of
\Planck\ detection.

We will address a topic on the purity of complete \Planck\ SZ sample and provide
a detailed study on galactic dust contamination
in a subsequent paper  by \citet{alej19}.

\section{Conclusions}
\label{sec:conclusions}

This paper opens a series of publications dedicated to the scientific
exploitation of the results obtained with the {\tt LP15} programme, a large,
systematic and uniform follow-up campaign of all the unconfirmed and new
\Planck\ PSZ2 sources in the northern hemisphere (i.e., having declinations
Dec.$>-15^{\circ}$ and not being included in the PSZ1 catalogue). With this
definition, our LP15 sample contains 190 objects (out of the 350 unconfirmed
objects in the full sky with no correspondence with the PSZ1 catalogue).

In this first paper, we summarize the information and discuss the results from
the first year of observations of LP15, where we carried out a detailed study of
106 targets. Each source has been carefully inspected, validated and classified
with a {\tt Flag} value, which was assigned according to our confirmation
criteria. We used the available spectroscopic and photometric redshifts, the
flux density profiles from the \Planck\ Compton $y$-map, dynamical properties of
the clusters and the optical richness. In particular, in this study we present a
refined procedure for the computation of the richness parameter, taking into
account the local background variance, and making it more robust in comparison
with other validation works.

Using our approach, out of the 106 objects, we have been able to confirm  41 new
\Planck\ PSZ2 sources. Among them, 31 were validated using  velocity dispersion calculated from our spectroscopic
redshifts.  We also presented updated redshift information for one cluster, PSZ2 G310.81+83.91, 
which is not included in the LP15 sample. Thus, in total we presented information for 
107 sources of the PSZ2 catalogue. In three cases, confirmed spectroscopically, we detected
the presence of multiple counterparts along the line of sight. The remaining 65
sources were classified as non-confirmed, either due to the absence of observed
over-densities (57), or due to a weak association with the SZ signal and
dissatisfaction with our validation requirements (8 objects with {\tt Flag}=3).
We looked for the possible explanations for this apparent absence of optical
counterparts.  We found, that most of the unconfirmed sources are associated
with areas of strong dust contamination with consequently possible spurious
enhancements of the SZ signal.  Also, most of our unconfirmed sources show low
values of neural network quality flag {\tt Q$\_$NEURAL}. We confirm that this
flag effectively separates between real and spurious SZ detections, and it
should be considered as a valuable tool during the validation process.

The work presented here contributes to a series of efforts to validate
completely all \Planck\ SZ sources, and ultimately will allow us to determine
the purity and efficiency of cluster detections in the \Planck\ SZ
catalogues.

\clearpage 
\begin{landscape}
\begin{table} 
\caption[]{\label{tab:newpsz2}List of 107 PSZ2 sources discussed in
  this paper. We show information for 106 LP15 cluster candidates plus one cluster from \citet{streb18}  with
  updated redshift information. The first four columns provide information from the PSZ2
  catalogue and correspond to the index number, official name of the SZ target, 
  the signal-to-noise ratio and neural network quality flag Q$\_$NEURAL respectively. The J2000 coordinates (column 5
  and 6) correspond to the BCG position or geometrical centre of the identified
  counterpart. Column 7 provides distance (in arcmin) between PSZ coordinates
  and optical centre. Column 8 shows the photometric redshift of the
  cluster. Columns 9, 10, and 11 list (if available) the mean spectroscopic
  redshift, spectroscopic redshift of the BCG (in the case of absence of the
  apparent BCG we write "-1"), and the number of galaxies with spectroscopic
  measurements. Columns 12 and 13 provide optical richness of the cluster and its
  significance, where possible (see Sec. \ref{sec:pho}) for details).  Column
  14 presents flag according to our validation criteria (Table \ref{tab:crit})
  . The last column includes comments.}
\label{tab:newpsz2}
\tiny
\centering
\begin{tabular}[h]{ccccccccccccccc}
    \hline \hline
ID & Planck name& S/N & Q$\_$NEURAL & R. A. & Decl. & distance & $z_{\rm phot}$ & $z_{\rm spec}$ & $z_{\rm spec}(BCG)$ & N$_{\rm spec}$  & R & $\sigma_R$ & flag & Comments \\
 \hline \hline
  34$^{d}$ & PSZ2 G009.04+31.09 &  5.04  & 0.98 & 16:18:26.70 & -04:11:11.06  & 1.52 & 0.25$^{b}$  & 0.246  & -1  &  34  &   31$^{a}$    &  - &   1 & WHL J161826.7-041111 \\  
  78    &   PSZ2 G023.87-13.88  &  4.90  & 0.95  & 19:25:40.06 & -14:11:25.04  & 1.52 & 0.12  & 0.0   &  0.0  &   0     &   30.0   &   2.64    &   2   &    fossil \\
  88    &   PSZ2 G027.23+15.73  &  4.89  & 0.01 &  -	         & -	    &  -	& -  &  -  &  -  &  0 &  0.0   &    0.0   &  ND  &	     \\
  90    &   PSZ2 G027.39+15.39  &  5.22  & 0.74 &  -	         &  -	    &  -	& -  &  -  &  -  &  0 &  0.0   &    0.0   &   ND  &	     \\
  91    &   PSZ2 G027.77+10.88  &  6.41  & 0.01 & -	         &  - 	    &  -	& -  &  -  &  -  &  0 &  0.0   &    0.0   &   ND   &	     \\
  106   &   PSZ2 G029.87-17.81  &  4.91  & 0.08 & -	         &  -	    &  -	& -  &  -  &  -  &  0 &  0.0   &    0.0   &   ND   &	     \\
  116   &   PSZ2 G032.77+19.69  &  4.78  & 0.99 & 17:40:23.30 & +08:41:12.12 &  2.10  & 0.39 & 0.369 & 0.368 &  61 &  24.6  &  3.28   &  1   &	   \\
  126$^{e}$ & PSZ2 G036.36+16.01 & 4.58  & 0.93 & 17:59:45.60 & +10:08:29.15 &  4.21 & 0.13 & 0.128 & -1  &  3  &  13$^{g}$  &  -   &  2   &	 \\
  127   &   PSZ2 G036.69-15.67  & 5.69   & 0.79 & -	         &  -	    &  -	& - &  - &  - &  0 &0.0   &    0.0    &  ND   &	\\  
  129   &   PSZ2 G036.77-16.51  & 4.62   & 0.03 & -	         &  -	    &  -	& - &  - &  - &  0 &0.0   &    0.0    &   ND   &	\\  
  130   &   PSZ2 G036.80-14.95  & 4.72   & 0.07 & -	         &  -	    &  -	& - &  - &  - &  0 &0.0   &    0.0    &   ND   &	\\ 
  133   &   PSZ2 G037.31-21.54  & 4.96   & 0.77 & 20:15:55.00  & -05:55:59.25 & 4.40  & 0.46 & 0.574 & -1 & 31 & 14.3  &   3.01   &   1   &	  \\ 
  143   &   PSZ2 G039.86+18.70  & 4.55   & 0.99 & -	         &  -	    &  -        & - &  -  &  - &  0 & 0.0   &    0.0    &   ND   &	  \\ 
  146-A$^{c}$ & PSZ2 G040.11-42.58 & 5.08  & 0.99 & 21:36:59.42 & -13:08:02.70 & 2.63 &0.17 & 0.202 & 0.201 & 54 & 45.7  &    7.8    &   1   &	   \\
  146-B$^{c}$ &              &             &  & 21:37:17.15 & -13:01:10.77 & 5.47 &0.57 & 0.62  & 0.62 & 15  & 4.9   &    1.71   &   1   &    \\
  155   &  PSZ2 G042.54+18.02 & 4.78  &   0.99 & 18:02:59.10  &  +16:21:15.37 & 6.56 &0.22 & 0.258 & -1 & 8	 & 40.2  &    3.28   &    3   &   \\
  158   &  PSZ2 G043.44-41.27 & 5.55  &  0.97  & 21:36:43.74  &  -10:19:01.69 & 1.23 &0.37 & 0.434 & -1 & 33  & 52.9  &    11.5   &   1   &   \\
  161$^{d}$ & PSZ2 G044.21+52.13 & 4.64 & 0.96 & 15:42:50.44 & +27:49:52.90  & 2.35  & 0.38 & 0.377 & -1   & 6  & 2.9   &  0.75   &   3   & WHL J154250.4+274953 \\  
  169   &    PSZ2 G045.20+15.63  & 4.65 & 0.91 & 18:16:04.76 & +17:47:00.12  & 1.30  & 0.13 & 0.126 & 0.128  & 22 & 16.2  &  1.67   &   1   &   \\
  171   &    PSZ2 G045.47+17.80  & 4.66 & 0.98  & 18:08:14.70 & +18:51:54.00  & 1.74  & 0.36 & 0.372 & 0.372  & 34 & 8.7   &  2.47   &   1   &   \\
  176$^{d}$ & PSZ2 G045.96-26.94 & 5.1  & 0.96 & 20:50:01.00 & -01:35:23.85  & 1.84  & 0.22 &  0.0  &  0.0   &  0 & 9.5   &  1.51    &   2   &  \\ 
  181   &  PSZ2 G046.39+11.71  & 4.62   &  0.96 & -	         &  -	    &  -  	& - &   - &   - &   0 & 	0.0   &    0.0    &   ND   &  \\
  193   &   PSZ2 G048.39-16.78 & 4.57   & 0.99 &  -	         &  -	    &  -  	& - &   - &   - &   0 & 	0.0   &    0.0    &   ND  &   \\
  208   &   PSZ2 G051.48-30.87 & 4.99   & 0.97 & -	         &  -	    &  -  	& - &   - &   - &   0 & 	0.0   &    0.0    &   ND  &   \\
  209   &   PSZ2 G052.08+46.13 & 4.75   & 0.94 &  -	         &  -	    &  -  	& - &   -  &  - &   0 & 	0.0   &    0.0    &   ND  &   \\
  217   &  PSZ2 G053.80+36.49  & 5.16   & 0.97 &  -	         &  -	    &  -  	& - &   -  &  -  &  0 & 	0.0   &    0.0    &   ND  &   \\
  233   &  PSZ2 G057.09+11.19  & 4.96   & 0.99 & 18:54:14.08 & +26:27:26.46 & 2.61  &0.35  & 0.377  &-1  & 20  &   14.3   &   1.49    &   1   &  \\
  248   &  PSZ2 G059.52+16.23  & 4.77   & 0.67 &  18:37:15.08 & +30:40:53.17 & 1.71  &0.24  & 0.284  &-1 &  17  &   100.8  &   31.64   &   1   &  \\ 
  250   &   PSZ2 G059.76+14.59 & 4.88   & 0.99 & 18:44:42.47 & +30:17:31.54 & 2.87  &0.28  & 0.303  &-1  & 9    &  -1   &   -1    &   1   &   \\
  266   &   PSZ2 G064.98+16.71 & 4.57   & 0.58 & 18:44:30.25 & +35:42:19.52 & 1.26  &0.16  & 0.222  &-1   &5    &  23.3   &   3.3	  &   1   &   \\ 
  269   &   PSZ2 G065.35-08.01 & 5.82   & 0.00 & -	         &  -	    &  -  &   -   & -  &  -  &  0   &   0.0    &   0.0	  &   ND   &    \\
   279$^{d}$ &   PSZ2 G066.59-58.51 & 5.01 & 0.91   &  23:07:11.30 & -07:31:43.15 & 3.58 & 0.36 &  0.334 & 0.334 & 16  &   2.1  & 0.61   &   3   & WHL J230711.3-073143   \\ 
  281    &   PSZ2 G066.76-08.42 & 4.81   & 0.92   &  -	         &  -	    &  -  &   -  &  -  &  -   & 0   &   0.0    &   0.0	  &   ND   &  \\ 
  284$^{d}$ &   PSZ2 G067.21-20.49 & 4.51 & 0.75  &  21:13:28.89 & +18:03:27.28 & 0.33 & 0.32  & 0.367  &0.366 & 14  &   -1   &   -1   &  1 & WHL J211328.9+180327  \\ 
  287   & PSZ2 G067.72-24.13   &  4.68  & 0.91 &  21:26:33.70 &  +16:04:47.30    &  2.37  & 0.10  &  0.0  & 0.0   &  0    &   -1   &  -1    &   3   &     \\ 
  295-A$^{c,d}$ &   PSZ2 G069.47-29.06 & 5.14 & 0.97  &  21:46:02.11 & +14:01:26.25 & 2.2  & 0.36  & 0.393  &0.393 & 44  &   9.2    &   1.92    &   1   &   \\
  295-B$^{c,d}$ &                      &    &    &  21:45:51.9  & +14:03:23.00  & 4.66  & 0.21  & 0.19  &-1 & 30  &   8.7    &  1.58   &   1   &   \\
  310   &   PSZ2 G072.96-12.26 & 4.58   & 0.84  &   -	         &  -	    &  -  &   -  & -   & -  &  0   &   0.0    &   0.0	  &   ND   &   \\
  319   &   PSZ2 G075.08+19.83 & 5.73   & 0.91 & 18:46:41.11 & +45:47:04.30  &1.27 & 0.2 &   0.0  &  0.0  &  0   &   18$^{g}$   &   -1    &   2   &   \\
  326   &   PSZ2 G076.51+21.73 & 4.68   & 0.99 & 18:38:35.82 & +47:33:35.78  &4.12 & 0.37  & 0.418 & 0.42  & 47   &  50.4   &   8.98    &   1   &   \\
  328   &   PSZ2 G076.81-32.57 & 4.68   & 0.92 & -	         &  -	    &  -  &   -  &  -   & -   & 0   &   0.0    &   0.0	  &   ND   &   \\
    \hline   
  \end{tabular}
\end{table}
\end{landscape}

 \addtocounter{table}{-1}
\begin{landscape}
\begin{table}
\caption{Continue.}
\tiny
\centering
\begin{tabular}[h]{ccccccccccccccc}
    \hline \hline
ID & Planck name& S/N & Q$\_$NEURAL & R. A. & Decl. & distance & $z_{\rm phot}$ & $z_{\rm spec}$ & $z_{\rm spec}(BCG)$ & N$_{\rm spec}$  & R & $\sigma_R$ & flag & Comments \\
 \hline \hline
 330$^{d}$   &   PSZ2 G077.67+30.59 & 5.03   & 0.99 & 17:46:50.85 & +50:31:12.28 & 3.18  &0.22  & 0.221 & 0.219 & 10  &   6.6    &   1.43    &   1   & WHL J174650.9+503112  \\
 337   &   PSZ2 G079.36+38.06 & 4.94 & 0.99  &  16:59:09.10 & +52:06:09.80 & 2.92  &0.26  & 0.299 & -1  & 53  &   31.8   &   11.45   &   1   &   \\
  351   &   PSZ2 G081.60+18.47 & 5.57  & 0.99 &  19:07:17.98 & +51:05:14.46 & 2.78 & 0.44  & 0.518 & 0.519  &21  &   11.4   &   2.57    &   1   &   \\
  353   &   PSZ2 G082.06+27.23 & 5.32 & 0.99  &  -	         &  -	    &  -  &    -  &  -  &  -  &  0    &  0.0    &   0.0	  &   ND   &   \\ 
  364   &   PSZ2 G084.15-08.63 & 4.82 & 0.00  &  -	         &  -	    &  -  &   -   & -  &  -   & 0   &   0.0    &   0.0	  &   ND   &   \\ 
  371$^{d}$ &   PSZ2 G084.69-58.60 & 4.73 & 0.99 & 23:36:37.50 & -01:27:52.30 &7.17  &0.2  & 0.185  &  -1 &  2    &  21.6   &   5.04    &   2   & WHL J233637.5-012752 \\  
  380   &   PSZ2 G086.07-41.99 & 4.78 & 0.99  &  -	         &  -	    &  -  &   - &  - &  - &  0   &  0.0    &   0.0  &   ND   &    \\
  382   &   PSZ2 G086.35-13.94 & 5.42  & 0.41  &  21:49:41.40 & +35:43:13.89 & 0.84 & 0.26  & 0.278   & 0.277  &  11    &  33.0   &   5.93    &   1   &    \\
   400   &   PSZ2 G089.06-11.79 & 5.71 & 0.85   &  21:52:58.09 & +39:04:30.53 & 1.14 & 0.38  & 0.455  &0.455 & 32   &  12.3   &   1.79    &   1   &   \\ 
  407   &   PSZ2 G090.12-13.87 & 4.86 & 0.99  &  22:03:21.46 & +38:03:39.42 & 4.08 & 0.1   & 0.071    &0.074  &  15   &   26.9   &   2.66    &   1   &   \\ 
  412   &   PSZ2 G091.56+08.50 & 4.77 & 0.99  &  20:36:18.61 & +54:59:04.90 & 0.39 & 0.26  & 0.27  & 0.27  & 11   &  86.7   &   20.82   &   1   &   \\ 
  415   &   PSZ2 G092.11-33.73 & 5.81 & 0.03   &  -	         &  -	    &  -  &   -   & -   & -  &  0   &   0.0    &   0.0	  &   ND   &   \\ 
  420   &   PSZ2 G092.64+20.78 & 5.12 & 0.92  &  19:16:45.42 & +61:40:41.47  &0.15  &0.38 &  0.545 & 0.549  &39  &   -1   &   -1    &   1   &   \\ 
  426   &   PSZ2 G093.71-30.90 & 5.0  & 0.00  &  -	         &  -	    &  -  &   -  &  - &   -  &  0   &   0.0    &   0.0	  &   ND   &   \\ 
  429   &   PSZ2 G093.94+13.75 & 4.88 & 0.00  &  -	         &  -	    &  -  &   -  &  -  &  -  &  0   &   0.0    &   0.0	  &   ND   &   \\ 
  444   &   PSZ2 G095.75-11.16 & 5.32 & 0.00  &  -	         &  -	    &  -  &   -  &  -  &  -  &  0   &   0.0    &   0.0	  &   ND  &   \\ 
  445   &   PSZ2 G096.10+12.46 & 5.08 & 0.05  &  -	         &  -	    &  -  &   -  &  -  &  -  &  0    &  0.0    &   0.0	  &   ND  &   \\ 
  447   &   PSZ2 G096.43-20.89 & 5.8  & 0.99  &  22:48:09.42 & +35:33:49.49 & 0.48 & 0.2   & 0.226  &  -1 &   34    &  33.3   &   4.31    &   1   &    \\
  454   &   PSZ2 G097.37-17.11 & 4.73 & 0.96  &  22:42:35.75 & +39:15:19.56  &2.12  &0.4   & 0.456 & 0.456 & 32   &  37.8   &   7.85    &   1   &    \\
  463   &   PSZ2 G098.39+57.68 & 5.07 & 0.86   &  -	         &  -	    &  -  &   -  &  -  &  -  &  0   &   0.0    &   0.0	  &   ND   &   \\ 
  465   &   PSZ2 G098.62+51.76 & 4.53 & 0.92  &  -	         &  -	    &  -  &   -  &  -  &  -  &  0  &    0.0    &   0.0	  &   ND  &   \\ 
  480   &   PSZ2 G100.07+17.06 & 5.01 & 0.00  &  -	         &  -	    &  -  &  -  &  -   & -   & 0   &   0.0    &   0.0	  &   ND  &   \\ 
  483$^{d}$   &   PSZ2 G100.22+33.81 & 5.68 & 0.99  &  17:13:41.57 & +69:21:45.24 & 0.65 & 0.61 &  0.598 & -1  & 18   &  3.3    &   1.27    &   1   &   \\ 
  484   &   PSZ2 G100.38+16.73 & 5.08 & 0.01  &  -	         &  -	    &  -  &   - &  -  &  -  &  0   &   0.0     &   0.0	  &   ND   &    \\
  485   &   PSZ2 G100.45+16.79 & 11.78 & 0.02  &  -	         &  -	    &  -  &   -  &  -  &  -  &  0   &   0.0     &   0.0	  &   ND  &   \\ 
  497   &   PSZ2 G104.15-38.85 & 6.41 & 0.00  &  -	         &  -	    &  -  &  -  &  - &   -   & 0   &   0.0    &   0.0	  &   ND  &    \\
  501   &   PSZ2 G104.58-15.41 & 4.62 & 0.00  &  -	         &  -	    &  -  &  -  &  - &   -   & 0   &   0.0    &   0.0	  &   ND  &    \\
  505   &   PSZ2 G105.00+39.68 & 4.97 & 0.91  &  15:52:52.42 & +70:30:57.64 & 1.70 & 0.2  &  0.2  &  0.201 & 30  &   31.7   &   6.18    &   1   &   \\
  512   &  PSZ2 G106.11+24.11 & 5.7   & 0.99  &  19:21:31.86 & +74:33:27.43 & 0.51 & 0.1  &  0.0  &  0.0   & 0   &   -1   &   -1    &   2   &  $z_{\rm phot}$=0.15 \citep{boada18} \\
  514   &  PSZ2 G106.21+26.32 & 4.9   & 0.99  &  18:48:31.20 & +75:03:29.99 & 0.09 & 0.08 &  0.0   & 0.0   & 0   &   -1   &   -1    &   2   &  1RXS J184828+750326 \\
  522   &  PSZ2 G107.41-09.57 & 10.68 & 0.01  &  -	         &  -	    &  -  &  -  &  -  &  -  &  0   &   0.0    &   0.0	  &   ND   &   \\
  525   &   PSZ2 G107.83-45.45 & 7.09 & 0.86  &  00:07:35.62   &  +16:07:01.87  & 0.83 & 0.55 &  0.567  &  0.567  & 2  &  6.8   &   2.7       &   2   &  $z_{\rm phot}$=0.55 \citep{boada18} \\
  538   &   PSZ2 G110.69-46.25 & 5.04 & 0.92  &  -	         &  -	    &  -  &  -  &  -  &  -  &  0   &   0.0    &   0.0	  &   ND   &   \\
  542   &   PSZ2 G112.07-39.86 & 5.72 & 0.01  &  -	         &  -	    &  -  &  -  &  -  &  -   & 0    &  0.0    &   0.0	  &  ND   &   \\
  568   &  PSZ2 G116.05+20.00 & 5.16  & 0.99  &  -	         &  -	    &  -  &  -   & -  &  -   & 0    &  0.0    &   0.0	  &   ND   &   \\
  574   &  PSZ2 G117.11+11.48 & 5.13  & 0.03  &  23:28:53.00 & +73:22:13.00 & 5.92 & 0.35  & 0.26  & -1  & 18   &  -1   &   -1    &   3   &   \\
  575   &  PSZ2 G117.38-52.47 & 5.45  & 0.49  &  -	         &  -	    &  -  &  -  &  -   & -   & 0   &   0.0    &   0.0	  &   ND  &   \\
  584   &   PSZ2 G118.79+47.50 & 5.18 & 0.97  &  13:24:21.08 & +69:17:24.72 & 7.46 & 0.34  & 0.0  &  0.0  &  0   &   43.5   &   11.05   &   2   &   \\
  591   &   PSZ2 G120.36+26.03 & 5.29 & 0.00  &  -	         &  -	    &  -  &  -   & -  &  -   & 0   &   0.0    &   0.0	  &   ND  &   \\
  593-A$^{c}$ & PSZ2 G120.76+44.14 & 5.58 & 0.99 &  13:12:53.57 & +72:55:06.22 & 2.02 & 0.28 &  0.296 & 0.295 & 41  &   17.0   &   5.68    &   1   &  Abell 1705 \\
  593-B$^{c}$ &              &           &   &  13:12:31.07 & +72:50:54.41 & 2.54 & 0.36 &  0.363 & -1  & 12   &  -1   &   -1    &   1   &  gravitational arc\\
  597   &   PSZ2 G121.87-45.97 & 4.85 & 0.01  &  -	         &  -	    &  -  & -  &  - &  -  &  0   &   0.0    &   0.0	  &   ND   &   \\
  609   &   PSZ2 G124.11+25.02 & 5.52 & 0.00  &  -	         &  -	    &  -  &  -   & -  &  -  &  0   &   0.0    &   0.0	  &   ND  &   \\
  612   &  PSZ2 G125.11+28.14 & 4.98  & 0.00  &  -	         &  -	    &  -  &  -  &  -  &  -  &  0   &   0.0    &   0.0	  &   ND  &   \\
  617   &   PSZ2 G125.55+32.72 & 6.48 & 1.00  &  11:25:46.87 & +83:55:04.58 & 2.43   & 0.27 &  0.0  &  0.0  &  0  &    6.5    &   2.12    &   2   &  $z_{\rm phot}$=0.20 \citep{boada18}  \\
  636   &  PSZ2 G128.15-24.71 & 4.74  & 0.94  &  01:15:25.70 &  +37:56:01.00 & 1.52 & 0.2 &   0.263 & -1  & 19 &	36.7  &    9.37   &   1   &   \\
 \hline   
  \end{tabular}
\end{table}
\end{landscape}

\addtocounter{table}{-1}
\begin{landscape}
\begin{table}
\caption{Continue.}
\begin{center}
\tiny
\centering
\begin{tabular}[h]{ccccccccccccccc}
    \hline \hline
ID & Planck name& S/N & Q$\_$NEURAL & R. A. & Decl. & distance & $z_{\rm phot}$ & $z_{\rm spec}$ & $z_{\rm spec}(BCG)$ & N$_{\rm spec}$  & R & $\sigma_R$ & flag & Comments \\
 \hline \hline
  655   &  PSZ2 G134.26-44.28 & 5.08  & 0.01  &  -	         &  -	    &  -  &  - &   -  &  -  &  0   &   0.0    &   0.0	  &   ND  &   \\
  666   &  PSZ2 G135.94-68.22 & 6.86  & 0.00  &  -	         &  -	    &  -  &  -  &  - &   -  &  0  &    0.0    &   0.0	  &   ND  &   \\
  668   &  PSZ2 G136.31+54.67 & 6.91  & 0.16  &  -	         &  -	    &  -  &  -  &  -  &  -  &  0   &   0.0    &   0.0	  &   ND  &   \\
  673$^{e}$    &   PSZ2 G137.24+53.93 & 7.87 & 0.08  &  11:40:59.55 & +61:07:07.04  &4.61  &0.48  & 0.476 & 0.477  &19  &   2.3    &   2.74  &  1  & WHL J114059.5+610707 \\  
  684   &   PSZ2 G139.72-17.13 & 5.11 & 0.98  &  02:19:44.22 & +42:50:13.26 & 2.06 & 0.12  & 0.155 & 0.156 & 18  &   9.1    &   1.2	  &   1   &  \\  
  705   &   PSZ2 G144.84-35.16 & 4.83 & 0.98  &  -	         &  -	    &  -  &  -  &  -  &  -  &  0    &  0.0    &   0.0	  &   ND  &   \\
 714$^{e}$   &   PSZ2 G146.16-48.92 & 5.1 & 0.00  & 01:52:41.75 & +11:13:01.36  &3.01 & 0.49 &  0.491 & 0.491  &20   &  0.3    &   0.13    &   3   & \\  
  723   &  PSZ2 G148.60-48.61 & 4.71 & 0.01   &  -	         &  -	    &  -  &  -   &  -   &  -   &  0    &   0.0    &   0.0	  &   ND   &  \\ 
  744   &  PSZ2 G153.56+36.82 & 15.89 & 0.00  &  -	         &  -	    &  -  &  -   &  -   &  -   &  0    &   0.0    &   0.0	  &   ND  &  \\ 
  763   &  PSZ2 G158.45-42.92 & 4.8  & 0.98   &  -	         &  -	    &  -  &  -   &  -   &  -   &  0    &   0.0    &   0.0     &  ND   &  \\ 
  771   &  PSZ2 G161.73-28.58 & 4.8   & 0.01  &  03:18:09.80 & +23:01:53.40 & 2.13  & 0.33   & 0.438   &0.442  & 12    &  9.0    &   2.6     &   3   &  \\ 
  776   &  PSZ2 G163.22-26.48 & 6.34 & 0.02   &  -	         &  -	    &  -  &  - &   - &   - &   0  &    0.0    &   0.0	  &   ND  &  \\ 
  785   &  PSZ2 G164.85-16.55 & 5.01  & 0.01  &  -	         &  -	    &  -  &  - &   - &   - &   0  &    0.0    &   0.0	  &   ND  &  \\ 
  795   &   PSZ2 G166.27-25.02 & 8.08 & 0.01  &  -	         &  -	    &  -  &  - &   - &   - &   0  &    0.0    &   0.0	  &   ND  &   \\
  796   &   PSZ2 G166.27-24.71 & 9.57 & 0.02  &  -	         &  -	    &  -  &  - &   - &   - &   0  &    0.0    &   0.0	  &   ND  &   \\
  801   &   PSZ2 G167.63-43.99 & 4.87 & 0.07  &  -	         &  -	    &  -  &  - &   - &   - &   0  &    0.0    &   0.0	  &   ND  &   \\
  813   &  PSZ2 G171.79-42.08 & 5.84  & 0.00  &  -	         &  -	    &  -  &  - &   - &   - &   0  &    0.0    &   0.0	  &   ND  &   \\
  827   &   PSZ2 G176.07-26.95 & 4.9  & 0.99  &  -	         &  -	    &  -  &  - &   - &   - &   0  &    0.0    &   0.0	  &   ND   &  \\ 
  845   &  PSZ2 G181.88-30.77 & 9.29  & 0.02  &  -	         &  -	    &  -  &  - &   - &   - &   0  &    0.0    &   0.0	  &   ND  &  \\ 
  1244$^{d}$ & PSZ2 G269.02+46.30 & 4.67 & 0.90 & 11:19:07.43 & -10:22:57.81 & 1.22  & 0.12$^{b}$  & 0.115  & 0.114  &  9  &   -1    &  -1	  &   3   & \\  
  1465$^{d,f}$  &  PSZ2 G310.81+83.91 & 8.28 & 0.01   &  12:55:18.02  & +21:02:31.22 & 5.0  & 0.46  & 0.429 &  -1  &  2   &   5.7   &   1.7	  &   2   & \\  
   1621  &  PSZ2 G347.96+80.46 & 4.73  & 0.93  &  -	        &  -	    &  -  & -  &  -   & -  &  0    &  0.0    &   0.0	  &   ND  &  \\ 
  1626  &  PSZ2 G349.18+38.66 & 5.15 & 0.99   &  15:11:40.98 & -11:11:27.36  &6.01 & 0.1  &  0.109 & 0.108  &26   &  -1   &   -1    &   1   &   \\
\hline   
 \end{tabular}
\end{center}
\small 
$^{a}$ Richness calculated using SDSS DR12 data data (no LP15 photometric data available)\\
$^{b}$ Photometric redshift obtained from SDSS DR12 data \citet{streb18} \\
$^{c}$ SZ targets identified with the ID followed by an A or B label indicate the presence of multiple counterparts\\
$^{d}$ Confirmed in \citet{streb18}. New LP15 photometric/spectroscopic data are available \\
$^{e}$ Clasified as "potentially associated" in \citet{streb18}\\
$^{f}$ An extra source from \citet{streb18} with new spectroscopic information. Not included in LP15 sample \\
$^{g}$ Richness calculated without local background subtraction due to the small FOV or poor observing conditions \\
\end{table}

\end{landscape}

\begin{acknowledgements}
This article is based on observations made with a) the Gran Telescopio Canarias
operated by the Instituto de Astrof\'{\i}sica de Canarias, b) the Isaac Newton
Telescope, and the William Herschel Telescope operated by the Isaac Newton Group
of Telescopes, and c) the Italian Telescopio Nazionale Galileo operated by the
Fundaci\'{o}n Galileo Galilei of the INAF (Istituto Nazionale di
Astrofisica). All these facilities are located at the Spanish Roque de los
Muchachos Observatory of the Instituto de Astrof\'{\i}sica de Canarias on the
island of La Palma. This research has been carried out with telescope time
awarded for the programme 128-MULTIPLE-16/15B. Also, during our analysis, we
used the following databases: the SZ-Cluster Database operated by the Integrated
Data and Operation Center (IDOC) at the IAS under contract with CNES and CNRS
and the Sloan Digital Sky Survey (SDSS) DR14 database. Funding for the SDSS has
been provided by the Alfred P. Sloan Foundation, the Participating Institutions,
the National Aeronautics and Space Administration, the National Science
Foundation, the U.S. Department of Energy, the Japanese Monbukagakusho, and the
Max Planck Society. This work has been partially funded by the Spanish Ministry
of Economy and Competitiveness (MINECO) under the projects ESP2013-48362-C2-1-P,
AYA2014-60438-P and AYA2017-84185-P. AS and RB acknowledge financial support
from the Spanish Ministry of Economy and Competitiveness (MINECO) under 2011
Severo Ochoa Programme MINECO SEV-2011-0187. HL is funded by PUT1627 and IUT26-2 grants from
the Estonian Research Council and by the European Structural Funds grant for the
Centre of Excellence "Dark Matter in (Astro)particle Physics and Cosmology"
TK133. Some of the results in this paper have been derived using the {\sc
  HEALPix} \cite{Healpix} package.

\end{acknowledgements}

\bibliographystyle{aa}


%



\end{document}